\documentclass[11pt]{article}
\usepackage[latin1]{inputenc}
\usepackage{times}
\usepackage{dsfont}
\usepackage{amsfonts}
\usepackage{float}
\usepackage{amsmath}
\usepackage{latexsym}
\usepackage{xr} 
\usepackage[numbers]{natbib}
\usepackage{color}
\usepackage{multirow}
\usepackage{makecell}
\usepackage{rotating}
\usepackage{url}
\usepackage{doi}

\relax 
\allowdisplaybreaks[3]

\numberwithin{equation}{section}
\DeclareMathOperator{\logit}{logit}

\newcommand{\red}[1]{{\textcolor{black}{#1}}}
\newcommand{\redd}[1]{{\textcolor{black}{#1}}}

\begin{document}

\title{\Large \bf Causal inference methods for small non-randomized studies: Methods and an application in COVID-19
}
\author{Sarah Friedrich and Tim Friede \\[1ex] 
{\small Department of Medical Statistics, University Medical Center G\"ottingen, Germany\footnote{Humboldtallee 32, 37073 G\"ottingen, Germany}}\\
{\small sarah.friedrich@med.uni-goettingen.de, tim.friede@med.uni-goettingen.de}
}
\maketitle

\begin{abstract} 
%(max. 250 words)
The usual development cycles are too slow for the development of vaccines, diagnostics and treatments in pandemics such as the ongoing SARS-CoV-2 pandemic.  Given the pressure in such a situation, there is a risk that findings of early clinical trials are overinterpreted despite their limitations in terms of size and design. Motivated by a non-randomized open-label study investigating the efficacy of hydroxychloroquine in patients with COVID-19, we describe in a unified fashion various alternative approaches to the analysis of non-randomized studies. % and apply them to the example study exploring the question whether different methods might have led to different conclusions. 
A widely used tool to reduce the impact of treatment-selection bias are so-called propensity score (PS) methods. Conditioning on the propensity score allows one to replicate the design of a randomized controlled trial, conditional on observed covariates. Extensions include \redd{the g-computation approach}, which is less frequently applied, in particular in clinical studies.
\redd{Moreover, doubly robust estimators provide additional advantages.}
 Here, we investigate the properties of propensity score based methods including \redd{three variations of doubly robust estimators} in small sample settings, typical for early trials, in a simulation study. %We conclude that \blue{the doubly robust g-computation has some desirable properties and should be more frequently applied in clinical research}. %In the hydroxychloroquine study, \red{the g-computation approaches} resulted in very wide confidence intervals indicating much uncertainty. We speculate that application of the method might have prevented some of the hype surrounding hydroxychloroquine in the early stages of the SARS-CoV-2 pandemic. 
R code for \redd{the} simulations is provided.  
\end{abstract}

\noindent{\bf Keywords:} COVID-19; Causal Inference; Propensity Score; Small samples

\vfill
\vfill

\newpage

\section{Introduction}\label{sec:int}
%(max. 500 words)

Pandemic situations such as the currently ongoing SARS-CoV-2 pandemic require the fast development of diagnostics, vaccines and treatments. As the usual development programs are too long in these situations, more efficient development pathways are sought. These include more innovative approaches such as platform trials and adaptive designs \citep{Stallard2020}. Furthermore, in situations of desperate medical need such as with COVID-19, early clinical trials might receive more attention than they would normally do. \redd{Ferreira et al.~call this a ``disruption of medical and scientific paradigms'' \citep{ferreira2020decline}.}
In March 2020, for instance, \red{Gautret et al.}~\cite{gautret2020hydroxychloroquine} published a report of a small open-label non-randomized controlled study suggesting that ``hydroxychloroquine treatment is significantly associated with viral load reduction/disappearance in COVID-19 patients''. Although typically not much notice would have been taken of such a small-scale study with its methodological limitations, the treatment was haled ``a game changer'' by the US president putting pressure on the regulatory authorities to license the drug for COVID-19 \citep{Rome2020}. 

In particular when a lot of importance is placed on non-randomized studies, their analyses and interpretation must be robust. Non-randomized studies might be prone to bias due to confounding.
\red{One common approach to deal with this is covariate adjustment in regression models. \redd{In clinical trial applications with a binary outcome, logistic regression is usually the method of choice.}
However, in the case of small sample sizes, the number of possible variables to adjust for is limited by the observed events.}
\redd{Moreover, the use of odds ratios is not without criticism in the literature \citep{bland2000odds,cook2002advanced,cummings2009relative,sonis} and it is often argued that the risk difference is of greater importance to clinical decision makers \citep{Austin2010}.}
	
\red{Besides covariate adjustment a wide range of methods were proposed to deal with confounding.} A widely used tool to reduce the impact of treatment-selection bias in observational data are so-called propensity score (PS) methods. The propensity score is defined as a participant's probability of receiving treatment given the observed covariates \citep{Rosenbaum1983,Rosenbaum1984}. Conditioning on the propensity score allows one to replicate the design of a randomized controlled trial, conditional on \emph{observed} covariates. Extensions include the g-computation \citep{Hernan2020,robins1986new}, which is less frequently applied, in particular in clinical studies. \redd{Moreover, doubly robust estimators have been proposed. Here, it is sufficient that either the outcome or the propensity score model is correctly specified. Hence, a doubly robust estimator} does not rely on correct specification of both models.    

\red{Gautret et al.~}\cite{gautret2020hydroxychloroquine} did not apply any of these methods for non-randomized studies, but analyzed the trial as if it was randomized. Here, we describe in a unified fashion various alternative approaches \redd{and explore in simulations whether different methods might have led to different conclusions}. New evidence has emerged in the meanwhile and we now know that hydroxychloroquine is not an appropriate therapy in COVID-19 \citep{cavalcanti2020,Sattui2020}. \redd{Thus}, we wonder whether a more appropriate analysis of the study by \red{Gautret et al.~}\cite{gautret2020hydroxychloroquine} could have prevented much of the hype and as a result might have saved some resources.

In the context of the analysis of clinical registries and routine data including electronic health records some of the methods described above have widely been applied and their characteristics explored in simulation studies. Given the applications, simulation experiments naturally considered large-scale data sets \citep{austin2014comparison}. 
\red{To the best of our knowledge, propensity score methods for small samples have received less attention in the literature so far. Recently, Pirracchio et al.~\cite{pirracchio2012evaluation} compared PS matching and weighting estimators in small sample populations and Andrillon et al.~\cite{Andrillon2020} investigated properties of different matching algorithms.}
Here, we investigate the properties of propensity score based methods including g-computation in small sample settings, typical for early trials, in a simulation study. %The doubly robust g-computation has some desirable properties as the simulations will demonstrate, but has so far not gained the attention deserved in clinical applications. 

The manuscript is organized as follows. In Section \ref{sec:ex} we introduce the study by \red{Gautret et al.~}\cite{gautret2020hydroxychloroquine}, which motivated our investigations, in more detail. In Section \ref{sec:methods} several approaches to the analysis of non-randomized trials are described. Their properties are assessed in a simulation study, in particular in the setting of small sample sizes, in Section \ref{sec:simu}. We close with a brief discussion of the findings and the limitations of our study (Section \ref{sec:dis}).

\section{Motivating example in COVID-19}\label{sec:ex}
Gautret et al.~\cite{gautret2020hydroxychloroquine} conducted an open-label non-randomized study investigating the efficacy and safety of hydroxychloroquine in addition to standard of care in comparison to standard of care alone. The patients in the hydroxychloroquine group were all from the coordinating centre whereas the controls were recruited from several centres including the coordinating centre. In the coordinating centre, however, only those patients refusing therapy with hydroxychloroquine were included as controls. A total of 36 patients were included in the analyses, 20 patients receiving hydroxychloroquine and 16 control patients. Out of the 20 patients on hydroxychloroquine, 6 patients received in addition also azithromycin. For the purpose of illustration, we only consider two treatment groups, i.e. with and without hydroxychloroquine. The primary outcome was virological clearance at Day 6 (with Day 0 being baseline). The individual participant data of the study are reported in Supplementary Table 1 of \cite{gautret2020hydroxychloroquine}. The variables included in the table include the patient's age, sex, clinical status (asymptomatic, upper respiratory infection or lower respiratory infection), duration of symptoms, and results of daily PCR testing for Days 0 to 6. \red{Gautret et al.~}\cite{gautret2020hydroxychloroquine} report virological cure at Day 6 for 14 out of 20 patients treated with hydroxychloroquine and for 2 out of 16 in the control group, resulting in a p-value of $0.001$ in an analysis not adjusted for any covariates. 

The study by \red{Gautret et al.~}\cite{gautret2020hydroxychloroquine} has \red{been subject to }criticism, mainly due to its limitations in design including the small sample size, choice of control patients, open label treatment and study discontinuations \citep{Alexander2020}. Although some preclinical data suggested potentially beneficial effects \citep{Cortegiani2020}, there were also some early warnings regarding some potentially harmful effects \citep{funck2020}. In the meanwhile, data from large-scale randomized controlled trials are available \red{demonstrating} that hydroxychloroquine is not suitable for postexposure prophylaxis for or the treatment of COVID-19 \citep{Boulware2020,cavalcanti2020, Horby2020}. The timeline of events is nicely depicted in Figure 1 of a review by \red{Sattui et al.~}\cite{Sattui2020}.

\section{Alternative analysis methods}\label{sec:methods}

\subsection{\redd{The choice of effect measure}}
\redd{We consider a binary outcome $Y$ as well as a binary treatment $A$ (1: experimental treatment, 0: control) and a vector of observed covariates $L$.
In clinical trial applications with a binary outcome $Y$ as in our motivating example, logistic regression is usually the method of choice. This method of analysis experienced a huge increase in the 1980s \citep{altman1991statistics} and is still very prevalent in clinical applications. The natural estimate obtained by a logistic regression is the odds ratio
\begin{eqnarray}\label{or}
	OR = \frac{P(Y=1| A=1)/P(Y=0 | A=1)}{P(Y=1| A=0)/P(Y=0 | A=0)},
\end{eqnarray}
i.~e.~the ratio of the odds of having the outcome under treatment and the odds of experiencing the outcome in the control group.
%Back in the 1980s, logistic regression was used more often than linear regression in medical literature \citep{altman1991statistics}. 
The use of odds ratios, however, is not without criticism in the literature, see e.~g.~\cite{bland2000odds,cook2002advanced,cummings2009relative,sonis}. Common arguments against the use of the OR include
that ORs are often not well understood by practitioners \citep{falagas2009well} or are misleadingly interpreted as relative risks \citep{bland2000odds}, which is only appropriate with rare events. 
Other possible effect measures include the risk ratio or the risk difference.% or the number needed to treat (NNT). 
It is often argued that the risk difference is of greater importance to clinical decision makers than relative effect measures such as the OR \citep{Austin2010}. Particularly in causal inference literature, there is a focus on the risk difference as effect measure. One reason for this is the issue of (non-)collapsibility: While marginal and conditional treatment effects coincide for the risk difference due to collapsibility, this is not true for the odds ratio \citep{Gail1984,Robinson1991,Sjoelander2016}. The same arguments also hold for the hazard ratio obtained from a Cox model in case of time-to-event data. Thus, additive models are the preferred choice here as well \citep{aalen,martinussen2013collapsibility}.
}

\subsection{Notation and some causal background}

In a randomized controlled trial, one would assume that due to randomization, the influence of the covariates $L$ is the same for treated and control patients. In observational studies, where allocation of the treatment is not in the hand
of the investigator, this direct comparison of the treatments may no longer be fair due to the influence of other
confounding factors, i.e., the distribution of the other risk factors $L$ may differ between treated and controls.
In order to imitate an RCT and to get valid estimates in this situation, \red{a common approach is} the so-called potential or counterfactual outcomes \red{framework} \citep{Hernan2020}:
Let $Y^{a=1}$ denote the outcome that would have been observed under treatment value $a=1$, and $Y^{a=0}$ the outcome that would have been observed under control ($a=0$). A causal effect is now defined as follows: we say that $A$ has a causal effect on $Y$ if $Y^{a=1} \neq Y^{a=0}$ for an individual. In practice, however, only one of these outcomes is observed for an individual under study. Therefore, we can only ever estimate an \emph{average causal effect}, which is present if $P(Y^{a=1}=1) \neq P(Y^{a=0}=1)$, i.~e.~the probability of the outcome under treatment is different from that under control, in the population of interest \citep{Hernan2020}.
%In our situation with a binary outcome and a binary treatment, we thus consider the causal odds ratio
%\begin{eqnarray}\label{causalOR}
%OR_c := \frac{P(Y^{a=1}=1)/(1-P(Y^{a=1}=1))}{P(Y^{a=0}=1)/(1-P(Y^{a=0}=1))},
%\end{eqnarray}
%as our primary outcome measure. Different approaches have been proposed to estimate $OR_c$ and we will discuss the most commonly used ones in the following. 
\redd{Thus, the causal risk difference is defined as 
	\begin{eqnarray}\label{causalOR}
RD_c = P(Y^{a=1}=1) - P(Y^{a=0}=1).		
	\end{eqnarray}
}

\subsection{Covariate adjustment of outcome model}

The conventional method to correct for baseline differences between groups is adjusting for all relevant patient characteristics in the outcome regression model. 
\redd{To many medical statisticians, the natural choice of model for binary outcome data would be a logistic regression model. This, however, gives an estimate of the odds ratio, not the risk difference we are interested in. 
Moreover, in the case of small sample sizes, the number of possible variables to adjust for is limited by the observed events. Otherwise, logistic regression estimators may be biased or the model may not converge due to separation, i.~e.~a single covariate or a combination of multiple covariates perfectly separates events from non-events \citep{altman2000we,steyerberg2019clinical,van2016no, van2019sample}. Possibilities to correct for this include Firth's penalized logistic regression and extensions thereof, see  \cite{puhr2017firth} and the references cited therein.
	To obtain the risk difference, one could use a generalized linear model with Binomial distribution and identity link function \citep{Austin2010}. However, the identity link function does not constrain the predicted probability to lie between 0 and 1 and the model often fails to converge \citep{Austin2010,cheung2007}. 
An alternative, which avoids convergence issues, is to use ordinary least-squares estimation (OLS) instead, i.~e.~we assume a linear relationship
$$
Y = \beta_0 + \beta_{trt} A + \alpha_1 \ell_{1} + \dots + \alpha_p \ell_{p}.
$$
Here, $\ell_1, \dots, \ell_p$ denote the observed values of the covariates $L_1, \dots, L_p$. 
Although OLS is usually used to analyze the mean of a continuous outcome, it can also be used to estimate risk differences, since the mean is equal to the risk in case of a binary response coded as 0 and 1 \citep{cheung2007}. 
Moreover, no distributional assumption is necessary to proof unbiasedness of the OLS estimator. In order to draw valid statistical inference, however, one has to consider robust variance estimators such as the Huber-White estimator. Since this is an asymptotic version of the robust variance, corrections are needed for small samples. The so-called HC3 variance estimator has been shown to perform best \citep{cheung2007}. The idea is to multiply the Huber-White robust variance by a correction factor that converges to 1 as sample sizes increase. The HC3 variance estimator is available in R (package sandwich) and SAS (PROC GLIMMIX).
}

\subsection{Propensity score based methods}
Several different methods have been proposed to estimate \redd{$RD_c$} in the literature, see e.~g.~\cite{Austin2011,Hernan2020} for an introduction. Many of these methods are based on the propensity score. The propensity score of individual $i$ is defined as $\widehat{p}_i:= \widehat{P}(A_i =1 |L_i)$, i.~e., the estimated probability of receiving treatment given the covariates. For all methods considered in this paper, we estimate the propensity score using a logistic regression model for treatment allocation based on all observed covariates, i.~e.
$$
P(A = 1|L) = \frac{\exp(\beta_0 + \beta_1 \ell_1 + \dots + \beta_{p} \ell_p)}{1+\exp(\beta_0 + \beta_1 \ell_1 + \dots + \beta_{p} \ell_p)}.
$$

In a practical data analysis, there are several possibilities for taking the propensity score into account. We will describe the most common methods in the following and apply them to the data example.

\paragraph{PS covariate adjustment}
%According to \cite{weitzen2004principles} and \cite{Shah_2005}, covariate adjustment using the propensity score was the most commonly used PS method in clinical literature. 
In this approach, the outcome $Y$ is regressed on the estimated propensity score $\widehat{p}$ and the treatment exposure $A$, %The estimated treatment effect $\widehat{OR}_c$ is the odds ratio for treatment exposure obtained from this logistic regression, 
\redd{i.~e., $
Y = \beta_0 + \beta_{trt} A + \beta_{ps} \widehat{p}$} and an estimator of the causal \redd{risk difference} is given by $\widehat{\beta}_{trt}$.

\paragraph{Matching on the propensity score}
Another possibility to balance treatment allocation is to match subjects on the propensity score. 
The idea is to find individuals with a similar propensity score in the treatment and the control group. There are various methods to match individuals. 
Particularly in small sample studies, it is impossible in practice to find exact matches. Thus, one needs to define an acceptable difference between the propensity scores of treated individuals and controls that will be used for matching. These differences are called \emph{calipers} and should be small enough to allow for ``a practical but meaningful equation of pairs'' \citep{Althauser1970}.
Following recently published recommendations \citep{Andrillon2020}, \red{where} propensity matching in small sample sizes \red{was investigated}, we performed a 1:1 nearest neighbor matching without replacement on the logit of the propensity score using calipers with a maximum width of 0.2 standard deviations. In this modification of classical nearest neighbor matching, subjects are only matched if the absolute difference of their propensity scores is within the pre-specified caliper distance \citep{austin2014comparison}. This distance is usually defined as a proportion of the standard deviation of the logit of the propensity score.
In R, this can e.~g.~be performed using the \emph{MatchIt}-package, where the PS-model, the method used for matching and the caliper can be specified. A caliper of 0.2 avoids matching dissimilar individuals. Note, however, that this setting differs from the default setting in R, where the caliper is set to 0. 

\redd{In a matched cohort, we can calculate the risk difference as 
$$
(\tilde{b}-\tilde{c})/n,
$$
where $\tilde{b}$ is the number of pairs where the treated subject experiences the event whereas the untreated subject does not, $\tilde{c}$ are the pairs where the untreated subject experiences the event but the treated does not and $n$ is the total number of matched pairs, see e.~g.~\cite{Austin2010} for details.
}

Note that since we match individuals without replacement, the matched data set will usually be smaller than the original study, sometimes even discarding treated individuals.

\paragraph{Inverse probability of treatment weighting (IPTW)}
Inverse probability weighting uses the whole data set, but weighs each individual with his or her (inverse) probability of receiving the actually given treatment. This way, it generates a pseudo-population with (almost) perfect covariate balance between treatment groups. More specifically, IPTW assigns weight $w_i=1/\widehat{p}_i$ to treated subjects and weight $w_i=1/(1-\widehat{p}_i)$ to controls. The resulting pseudo-population is analyzed using weighted regression with robust standard errors, \redd{which can, e.~g.~be obtained from the \emph{survey}-package in R}. 

\subsection{g-computation}
The fourth possibility to account for covariate unbalance that we consider is \emph{g-computation} \citep{Hernan2020,robins1986new}, also known as \emph{the parametric g-formula} or \emph{direct standardization}\red{, see \cite{snowden2011implementation} for an excellent introduction}. The idea is that the marginal counterfactual risk \red{can be written as}
$$
P(Y^{a}=1) = \sum_{\ell} P(Y^{a}=1| L = \ell)P(L=\ell) = \sum_{\ell} P(Y=1| L = \ell, A=a)P(L=\ell).
$$
Here, the sum is over all values $\ell$ of the confounder(s) $L$ that occur in the population. The right-hand side of this equation can now be estimated using the available data on $Y, A$ and $L$.
 %This method is based on the counterfactual outcomes, i.~e., the outcome a person would have had, had he/she received the treatment contrary to the one he/she actually received.
More precisely, we have to predict the \red{potential} outcome for every person $i$ in the population assuming 
\begin{enumerate}
	\item $i$ was treated
	\item $i$ was a control
\end{enumerate}
irrespective of the treatment actually received. \red{In order to achieve this}, we first fit a so-called Q-model to the data relating the outcome $Y$ to the exposure $A$ and to confounders $L$. For a binary outcome as in our situation, this is usually a logistic regression model. \red{Instead of using this model for estimation of the treatment effect, however, }we \red{use it to} predict $\widehat{P}(Y=1| L = \ell, A=1)$ and $\widehat{P}(Y=1| L = \ell, A=0)$ for all individuals by artificially creating two new data sets: One where $A=1$ for all individuals and one where $A=0$ for all individuals, respectively. \red{Thus, this step can be thought of as imputing the missing potential outcomes for each subject in the population. Finally, t}he causal \redd{risk difference $\widehat{RD}_c$} can be estimated \red{by averaging over the estimated probabilities of the outcome under treatment and control and applying} Equation~\eqref{causalOR}.

Confidence intervals for g-computation are usually obtained by a nonparametric bootstrap approach \red{\citep{efron1986bootstrap,Hernan2020}}, i.~e.~by drawing with replacement from the data and analyzing each bootstrap data set like we analyzed the original data. \redd{Resampling approaches like this lead} to asymptotically valid inference procedures \citep{efron1986bootstrap} and \redd{have} been shown to \redd{be} superior in small samples \redd{in various situations \citep{Beyersmann2013,bluhmki2018,Friedrich2017wild,friedrichMATS,Kon:2015,PBK}}.
\red{T}he number of bootstrap repetitions should be chosen reasonably large. We used 1,000 bootstrap repetitions in the simulation study\redd{, but recommend a higher number in real-life applications}.
Upper and lower 95\% confidence intervals are obtained using the 2.5 and 97.5 percentiles of the bootstrap distribution. 
Note that a statistical test can be obtained similarly by calculating the test statistic in each bootstrap sample and then comparing the original test statistic to the empirical $(1-\alpha)$-quantile of the bootstrap distribution. A $p$-value is obtained by counting how often the original test statistic is smaller than the bootstrap statistic and dividing this number by the conducted bootstrap replications\red{, see e.~g.~\cite{Friedrich2017wild,friedrichMATS} for similar approaches. To investigate \redd{the small sample performance of the bootstrap for g-computation in detail and determine} whether more elaborate bootstrap techniques might lead to better performance is part of future research.}

\subsection{Doubly robust estimators}

While IP weighting requires the propensity model to be correct, i.~e.~a correct model for the treatment $A$ conditional on confounders $L$, the g-formula requires a correct model for the outcome $Y$ conditional on treatment $A$ and the confounders $L$, the Q-model. A doubly robust \redd{(DR)} estimator, in contrast, is consistent if at least one of the two models is correctly specified. There are many types of doubly robust estimators \red{(see e.~g.~\cite{bang2005doubly,kang2007demystifying,van2010collaborative} and the references cited therein for an overview)}. \redd{We will focus on three different ones here. The first two are applied to the g-computation whereas the third is an extension of IPTW.}

\subsubsection*{\redd{Simple DR g-computation}}
\redd{The first DR estimator we consider is a very simple one} \citep{bang2005doubly,Hernan2020}: First, we estimate the weights $w_i$ as described above. We then fit our Q-model to the data including an additional covariate $z$, where  $z_i= w_i$ if $A_i=1$ and $z_i = -w_i$ if $A_i=0$. Finally, we again obtain a causal \redd{risk difference} from Equation~\eqref{causalOR}.
\red{This method is referred to as "Simple DR g-computation" in the following.}
\redd{Kang and Schafer \cite{kang2007demystifying} studied the performance of different DR estimators with a particular focus on the situation, where both the outcome and the PS model are misspecified. They found that this estimator behaves poorly, when the PS-model is misspecified and even state that "[t]he performance of this method is disastrous when some of the estimated propensities are small" \citep{kang2007demystifying}.}

\subsubsection*{\redd{DR using quintiles}}
\red{Another possibility for a DR estimator \redd{also studied by Kang and Schafer and found to "[perform] better than any of the other DR methods when the [models] are both incorrect" \citep{kang2007demystifying}}
		 is obtained by coarsening the logit of the estimated propensity score into five categories according to the quintiles. Thus, we include four dummy variables distinguishing among these categories in the Q-model for the g-computation, see \cite{kang2007demystifying} for details. We denote this approach "DR using quintiles" in the following.}

\subsubsection*{\redd{Augmented IPW (AIPW)}}
\redd{Another approach is to augment the IPTW estimator described above with a regression model for the outcome variable. 
	Thus, a separate outcome model of $Y$ on the confounders $L$ is needed. Details on the method can be found in \cite{goetghebeur2020,mao2019propensity} and the resulting AIPW estimator is implemented in the R package PSW. Note that this approach is closely connected to the simple DR g-computation described above: Bang and Robins \cite{bang2005doubly} found that the augmented estimator can be viewed as an unweighted regression including the inverse of the PS as a covariate \citep{goetghebeur2020}.}

\section{Simulation study}\label{sec:simu}
The set-up of our simulation study closely followed \red{Austin}~\cite{Austin_2007,Austin2010}. The data-generating process is as follows:
First, we generate $n$ covariates $x_1, \dots, x_n$ \redd{(see the following subsections for details)}. We then generate the treatment status for each subject $i=1, \dots N$ according to the model
\begin{eqnarray}\label{trtstatus}
\logit(p_{i, \text{treatment}}) = \beta_0 + \beta_1 x_1 + \dots + \beta_n x_n.
\end{eqnarray}

Treatment is then randomly assigned to each subject following a Bernoulli distribution with subject-specific probability of treatment assignment $A_i \sim \text{Bernoulli}(p_{i, \text{treatment}})$.
Next, the outcome $Y_i$ of each subject is simulated conditional on treatment assignment $A_i$ and the covariates associated with the outcome according to
\begin{eqnarray}\label{outcome}
\logit(p_{i, \text{outcome}}) = \alpha_0 + \beta_{trt} A_i + \alpha_1 x_1 + \dots + \alpha_n x_n
\end{eqnarray}
and $Y_i \sim \text{Bernoulli}(p_{i, \text{outcome}})$.
Here, $\beta_{trt}$ denotes the \redd{log-odds ratio relating treatment to the outcome. Thus, a value of $\beta_{trt}=0$ corresponds to the null effect, i.~e.~an odds ratio of 1 and a risk difference of 0.}

\redd{In contrast to odds ratios, the risk difference is collapsible, i.~e.~the average subject-specific risk difference is equal to the marginal risk difference. Based on 1,000 data sets of size $N=10,000$, we used the following procedure to determine the average risk difference and adjust the value of $\beta_{trt}$ to obtain the desired non-null risk differences:
For a fixed value of $\beta_{trt}$, we generate the counterfactual outcomes under treatment ($A_i =1$) and control ($A_i=0$) for each individual and calculate the marginal probabilities under treatment and control. The risk difference is then equal to the difference between these two marginal probabilities \citep{Austin2010}.
} Using an iterative process, we modified $\beta_{trt}$ until we got close enough to the desired marginal \redd{risk difference}.

Concerning the covariates, we considered three different scenarios:

\subsection{Scenario 1: The COVID-19 example}
The first scenario aimed at mimicking the data example. Thus, we generated four covariates: 
\begin{enumerate}
	\item $x_1$ (representing sex) followed a Bernoulli distribution with parameter 0.5
	\item $x_2$ (representing age) was drawn from a $N(45, 15)$ distribution and rounded to integers
	\item $x_3$ (clinical status) was simulated as a categorical covariate with three categories, i.~e.~a $\text{Bin}(2, 0.5)$ distribution
	\item $x_4$ (time since onset of disease) was generated from a uniform distribution on $[0, 10]$ and rounded to integers.
\end{enumerate}

Treatment status was then generated according to Equation~\eqref{trtstatus} with
$$
(\beta_0, \beta_1, \beta_2, \beta_{3,1}, \beta_{3,2}, \beta_4) = (-2.3, 0.31, 0.03, 1.099, -0.1054, 0.1031).
$$
Here, $\beta_{3,1}$ and $\beta_{3,2}$ correspond to the dummy-coded categories $x_{3, 1}$ and $x_{3, 2}$ for $x_3=1$ and $x_3=2$, respectively. The parameters were obtained from the data by univariate logistic regression. Note that this implies a moderate association of treatment with $x_1, x_{3,2}$ and $x_4$, a weak association with $x_2$ and a strong association with $x_{3,1}$.

Similarly, the outcome was generated following Equation~\eqref{outcome} with
$$
(\alpha_0, \alpha_1, \alpha_2, \alpha_{3, 1}, \alpha_{3,2}, \alpha_4) = (-1.06, 0.619, 0.0077, 0.9461, -1.3499, 0.0896),
$$
implying a moderate association with $x_1$ and $x_4$, a strong association with $x_3$ and a weak association with $x_2$.
The parameter $\beta_{trt}$ was varied to generate different \redd{risk differences} in the following way: For $\beta_{trt}=0$, \redd{the risk difference is equal to 0. For $\beta_{trt}=0.8678$ we get a risk difference of 0.16 and for $\beta_{trt}=3.128$ the true risk difference equals 0.4.}
Finally, $\beta_0=-2.3$ resulted in a similar distribution of treated individuals and controls as in the original data, yielding an average of $55.21\%$ of individuals in the treatment group. To study the influence of more or less unbalanced treatment groups, we also varied this parameter in the simulations. In particular, we additionally considered a treatment allocation of approx.~2:1 and 4:1.

\subsection{Scenario 2: Unmeasured confounder}
The parameters in this setting are identical to Scenario 1, but we additionally added an unmeasured confounder. Thus, we simulated a covariate $x_5$ following an $N(0, 1)$ distribution with a strong effect on both treatment assignment and outcome. Therefore, $\beta_5$ and $\alpha_5$ were set to $\log(5)$. However, $x_5$ entered neither the propensity score model nor the Q-model for the g-computation.
For a \redd{risk difference of 0.16 and 0.4, $\beta_{trt}$ was set to 1.1111 and 3.71, respectively.}

\subsection{Scenario 3: Following Austin's design}
This scenario \redd{is based on} \red{Austin}~\cite{Austin_2007}. Therefore, we used the same set-up as he did, namely simulating 9 binary covariates with different association to treatment assignment and outcome as described in Table \ref{tab:matrix}.

\begin{table}[ht]
	\centering
	\caption{Association to treatment assignment and outcome used in the simulation scenario motivated by \red{Austin}~\cite{Austin_2007} (Scenario 3).}
	\label{tab:matrix}
	\begin{tabular}{c|ccc}
		& \makecell{ Strongly associated \\ with treatment} & \makecell{Moderately associated \\ with treatment} & \makecell{Not associated \\ with treatment}\\\hline
		\makecell{Strongly associated\\ with outcome} & $x_1$ & $x_2$ & $x_3$\\
		\makecell{Moderately associated \\ with outcome} & $x_4$ & $x_5$ & $x_6$\\
		\makecell{Not associated \\ with outcome} & $x_7$ & $x_8$ & $x_9$\\
	\end{tabular}
\end{table}

Here, a strong association is represented by a coefficient of $\log(5)$, i.~e.~$\beta_1 = \beta_4 = \beta_7 = \alpha_1 = \alpha_2 = \alpha_3 = \log(5)$, while a moderate association has a coefficient of $\log(2)$, i.e.~$\beta_2 = \beta_5 = \beta_8 = \alpha_4 = \alpha_5 = \alpha_6 = \log(2)$.
We chose $\beta_0 =-3.5$ to obtain a balanced design with respect to treatment and $\alpha_0$ was set to $-5$.
 For more details on the simulation set-up, see \cite{Austin_2007}.
%. More precisely, the parameters were:
%$$
%(\beta_0, \beta_1, \beta_2, \beta_3, \beta_{4}, \beta_{5}, \beta_6, \beta_7, \beta_8, \beta_9) = (-3.5, \log(5), \log(2),0, \log(5), \log(2), 0, \log(5), \log(2), 0)
%$$
%and
%$$
%(\alpha_0, \alpha_1, \alpha_2, \alpha_3, \alpha_{4}, \alpha_{5}, \alpha_6, \alpha_7, \alpha_8, \alpha_9) = (-5, \log(5), \log(5),\log(5), \log(2), \log(2), \log(2), 0, 0, 0).
%$$
The propensity score model and the Q-model included all 9 covariates.
For a \redd{risk difference of 0.16 and 0.4, $\beta_{trt}$ was set to 1.032 and 2.448}, respectively.
In addition to Austin's setting with an equal treatment allocation of 1:1, we also considered a situation with approx.~4:1 treated patients.

An overview of all simulated scenarios is given in Table~\ref{tab:overview}.

\begin{table}[ht]
	\centering
	\caption{\redd{Overview of the simulated scenarios and where to find the results.}}
	\label{tab:overview}
	\begin{tabular}{c|cc||ccc}
		& $\beta_{trt}$ & \redd{true RD} & $\beta_0$ & \makecell{Percent treated\\ on average} & \makecell{simulated \\for true RD} \\ \hline
		& 0 & 0  &$-2.3$ & 55\% & 0, 0.16, 0.4 \\
		Scenario 1	& 0.8678 & 0.16 & $-1.8$ & 66\% & 0 \\
		&3.128 & 0.4    &$-1$ & 80\% & 0   \\ \hline
		& 0 & 0 &  & &\\
		Scenario 2  & 1.1111 & 0.16   & $-2.3$ & 54\% & 0, 0.16, 0.4 \\
		& 3.71 & 0.4 &   & & \\ \hline
		& 0 & 0   & $-3.5$ & 49\% & 0, 0.16, 0.4 \\
		Scenario 3 & 1.032 & 0.16 & $-1.5$& 80\% & 0 \\
		& 2.448 & 0.4  & && \\ \hline
		Results: & \multicolumn{2}{c||}{ Figures~\ref{fig:bias} and \ref{fig:cov}, Tables~\ref{tab:Scen1}--\ref{tab:Scen3}}& \multicolumn{3}{c}{Figures~\ref{fig:Scen1_unb} and \ref{fig:Scen3_unb}, Table~\ref{S1}}
	\end{tabular}
\end{table}

\redd{In order to compare our results for the risk difference to the approach of a logistic regression, i.~e.~to estimating a causal odds ratio, we have also performed our simulations for the odds ratio. The results are included in the supplemental material.}

\subsection{Simulation results}

To study the influence of small sample sizes on the methods, we simulated $N=40, 100, 1000$ individuals for each scenario. Simulations were performed in R Version 3.6.3 with 2,000 simulation runs and the bootstrap confidence intervals for the g-computation are based on 1,000 bootstrap replications. Note that while 1,000 bootstrap replications suffice in simulations, we would recommend a higher number, say 10,000, in real-life applications.

\begin{figure}[ht]
	\includegraphics[width = \textwidth]{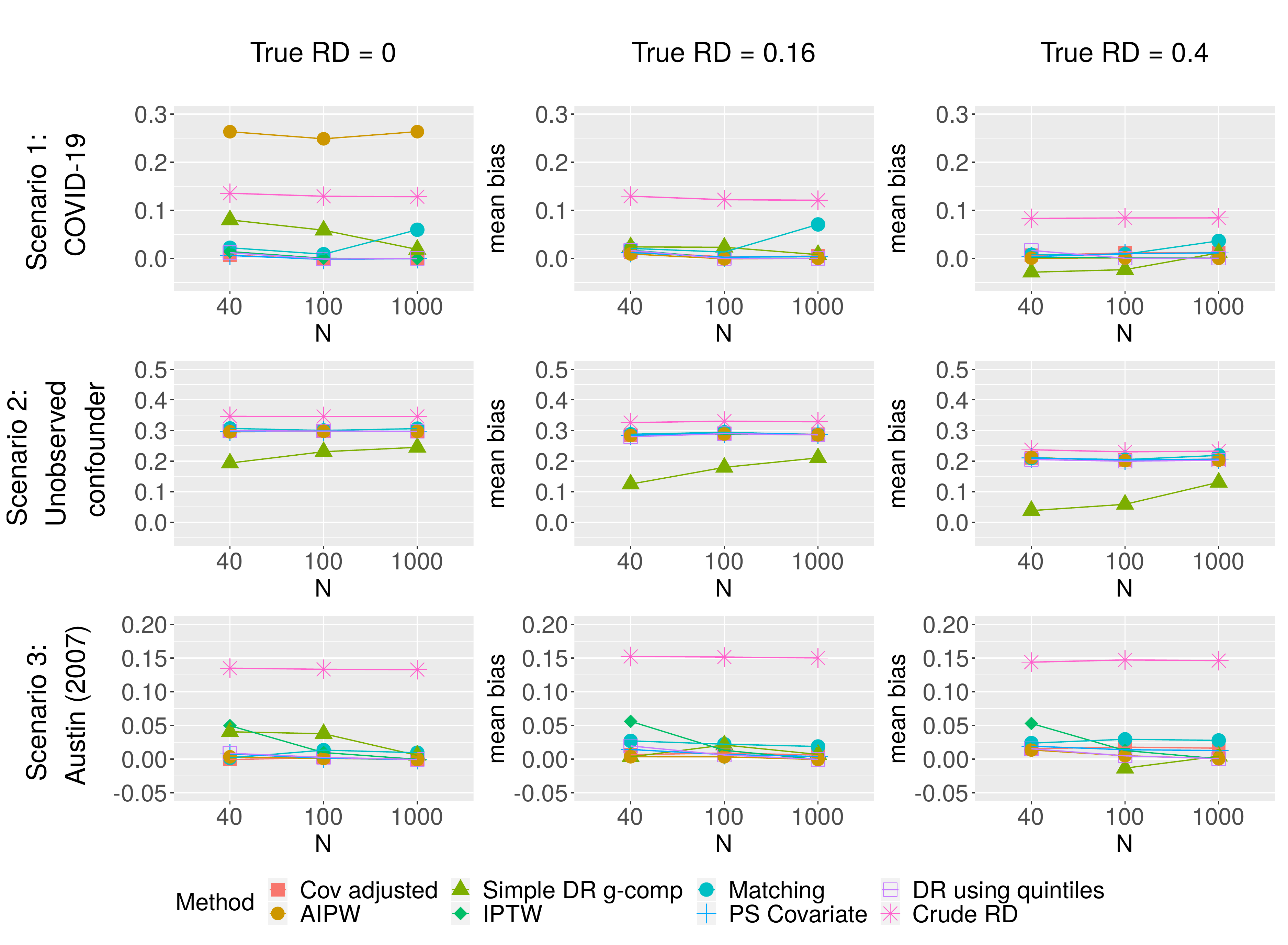}
	\caption{Displayed is the mean bias for the three scenarios (rows) and the three simulated \redd{risk differences} (columns).}
	\label{fig:bias}
\end{figure}

We used different measures to compare the results. With respect to the point estimators, we considered the mean bias, i.~e.~the mean difference between the true \redd{risk difference $RD$ and the estimated risk difference $\widehat{RD}$}. The results are displayed in Figure~\ref{fig:bias}. Moreover, the \red{root} mean square error of each estimated \redd{risk difference} (RMSE) \redd{and the median of the absolute errors (MAE), i.~e.~the median of $|\widehat{RD} - RD|$ are} displayed in Tables~\ref{tab:Scen1}--\ref{tab:Scen3}.

\begin{figure}[ht]
	\includegraphics[width = \textwidth]{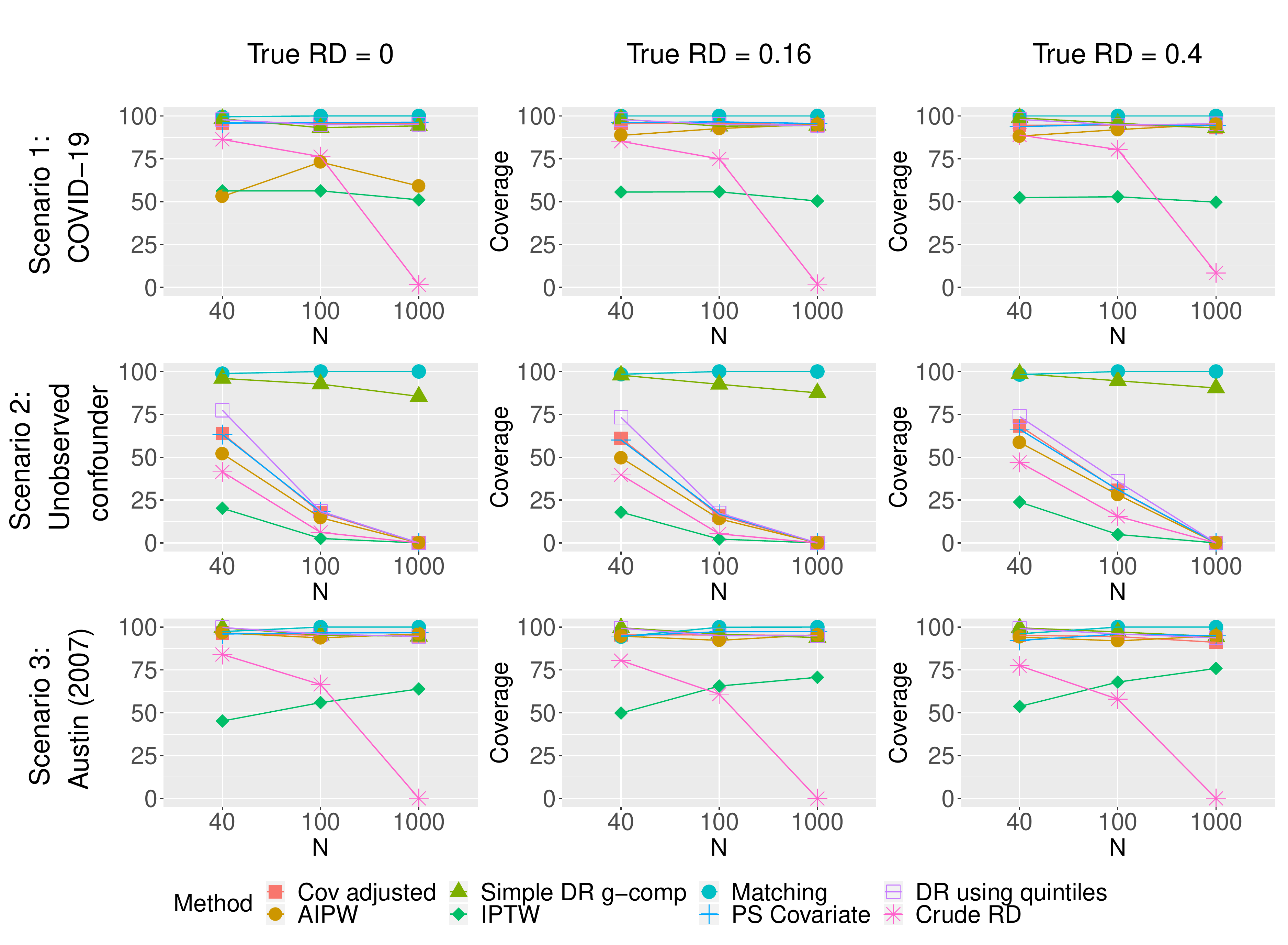}
\caption{Displayed is the coverage probability (in \%) for the three scenarios (rows) and the three simulated \redd{risk differences} (columns).}
	\label{fig:cov}
\end{figure}

Concerning the confidence intervals, we considered the percentage of 95\% confidence intervals that contained the true \redd{risk difference} (coverage probability) as well as the median length of the 95\% confidence interval. These measures are displayed in Figure~\ref{fig:cov} and  Tables~\ref{tab:Scen1}--\ref{tab:Scen3}, respectively. 
Finally, we also reported how often the \redd{methods failed, e.~g.~since no matching could be performed or the model did not converge}. These were excluded from the calculations and reported as failures in the tables.

For comparison, we included the crude \redd{as well as the covariate-adjusted risk difference}.

%The results are presented in Tables~\ref{tab:1}--\ref{tab:3_10}. Additional tables for the situation with increased unbalance between treatment groups (Scenario 1 and 3) as well as the results for a true OR of 2 in Scenario 2 and 3 have been moved to the supplement, see Tables~\ref{A-tab:A1}--\ref{A-tab:A3}.

% latex table generated in R 3.6.3 by xtable 1.8-4 package
% Fri Oct  2 10:15:38 2020
\begin{sidewaystable}[ht]
	\centering
	\caption{\redd{Median length of the 95\% confidence intervals (Length CI), root mean square error of the estimated treatment effect (RMSE), median absolute error (MAE) and number of models failed for Scenario 1: COVID-19. Crude RD is estimated by a linear regression adjusting only for treatment assignment, Cov adjustment additionally adjusts for baseline covariates, PS covariate denotes the method including the PS in the outcome regression model and Simple DR g-comp and DR using quintiles refer to the doubly robust g-computation methods, respectively}.}
	\label{tab:Scen1}
	\begin{tabular}{ccccccccccc}
		\hline
		 true RD & $N$ &  & Crude RD & \makecell{covariate \\ adjusted} & PS cov & matching & IPTW & \makecell{Simple DR \\g-comp} & \makecell{\red{DR using} \\ \red{quintiles}}  & AIPW \\ 
		\hline
		\multirow{4}{*}{0} & \multirow{4}{*}{$N=40$}  & CI Length & 0.6442 & 0.7157 & 0.736 & 4.162 & 0.2569 & 1.15 & 0.8127 & 0.5355 \\ 
		& & RMSE & 0.2105 & 0.169 & 0.1722 & 0.2185 & 0.1847 & 0.39 & 0.1886 & 0.2662 \\ 
		& & MAE & 0.15 & 0.1133 & 0.1142 & 0.1538 & 0.1246 & 0.3695 & 0.1265 & 0.2656 \\ 
		&  & Failures & 0 & 0 & 0 & 1404 & 0 & 1 & 0 & 0 \\ 
		\hline
		& \multirow{4}{*}{$N=100$} & CI Length & 0.397 & 0.4072 & 0.4251 & 4.219 & 0.1463 & 1.001 & 0.4006 & 0.5371 \\ 
		& & RMSE & 0.1621 & 0.09911 & 0.09948 & 0.1206 & 0.103 & 0.3388 & 0.1001 & 0.2497 \\ 
		& & MAE & 0.1319 & 0.06541 & 0.06507 & 0.08696 & 0.06772 & 0.2904 & 0.06678 & 0.2493 \\ 
		& & Failures & 0 & 0 & 0 & 1647 & 0 & 0 & 0 & 0 \\ 
		\hline
		& \multirow{4}{*}{$N=1000$} & CI Length & 0.1236 & 0.121 & 0.1265 & 4.099 & 0.04152 & 0.5677 & 0.1179 & 0.5354 \\ 
		 & & RMSE & 0.1319 & 0.0304 & 0.03046 & 0.06247 & 0.03058 & 0.1609 & 0.03002 & 0.2636 \\ 
		& & MAE & 0.1278 & 0.02041 & 0.02068 & 0.0596 & 0.02063 & 0.1054 & 0.02028 & 0.2632 \\ 
		& & Failures & 0 & 0 & 0 & 1998 & 0 & 0 & 0 & 0 \\ 
		\hline
		\hline
		
		\multirow{4}{*}{0.16} & \multirow{4}{*}{$N=40$} & CI Length & 0.6107 & 0.6981 & 0.7196 & 4.655 & 0.247 & 1.15 & 0.8185 & 0.5656 \\ 
		 &  & RMSE & 0.1986 & 0.1638 & 0.1665 & 0.2076 & 0.1783 & 0.3857 & 0.1851 & 0.1819 \\ 
		&   & MAE & 0.14 & 0.1113 & 0.113 & 0.1538 & 0.1219 & 0.375 & 0.1271 & 0.1228 \\ 
		 &  & Failures & 0 & 0 & 0 & 1404 & 0 & 0 & 0 & 58 \\ 
		 \hline
		 
		&\multirow{4}{*}{$N=100$} & CI Length & 0.3772 & 0.3988 & 0.4164 & 4.773 & 0.1431 & 0.9995 & 0.3957 & 0.3652 \\ 
		 & & RMSE & 0.1534 & 0.09643 & 0.09675 & 0.1154 & 0.1013 & 0.3292 & 0.09841 & 0.0999 \\ 
		 & & MAE & 0.1265 & 0.06524 & 0.06522 & 0.07529 & 0.0679 & 0.3021 & 0.06641 & 0.06872 \\ 
		 & & Failures & 0 & 0 & 0 & 1647 & 0 & 0 & 0 & 0 \\ 
		\hline
		 &\multirow{4}{*}{$N=1000$} & CI Length & 0.1174 & 0.1187 & 0.1245 & 4.651 & 0.04004 & 0.5618 & 0.1162 & 0.1181 \\ 
		 & & RMSE & 0.1245 & 0.03063 & 0.03058 & 0.07065 & 0.03059 & 0.1592 & 0.02999 & 0.03044 \\ 
		 & & MAE & 0.1213 & 0.02041 & 0.02066 & 0.07063 & 0.0201 & 0.1128 & 0.01986 & 0.02003 \\ 
		& & Failures & 0 & 0 & 0 & 1998 & 0 & 0 & 0 & 0 \\ 
		\hline
		\hline
		 \multirow{4}{*}{0.4} & \multirow{4}{*}{$N=40$} & CI Length & 0.5111 & 0.6016 & 0.6038 & 5.227 & 0.1897 & 1.15 & 0.75 &  0.487 \\ 
		 & & RMSE & 0.1509 & 0.1453 & 0.148 & 0.1679 & 0.159 & 0.379 & 0.1671 & 0.1594 \\ 
		 & & MAE & 0.1092 & 0.09989 & 0.1012 & 0.1 & 0.1082 & 0.325 & 0.1092 & 0.1007 \\ 
		 & & Failures & 0 & 0 & 0 & 1366 & 0 & 0 & 0 & 61 \\ 
		\hline
		 &\multirow{4}{*}{$N=100$} & CI Length & 0.3169 & 0.3422 & 0.3493 & 5.115 & 0.1158 & 0.9894 & 0.351 & 0.3153 \\ 
		 & & RMSE & 0.1177 & 0.08743 & 0.08763 & 0.09178 & 0.0911 & 0.3206 & 0.09038 & 0.09018 \\ 
		 & & MAE & 0.09127 & 0.05963 & 0.05935 & 0.06429 & 0.06132 & 0.266 & 0.05985 & 0.06143 \\ 
		 & & Failures & 0 & 0 & 0 & 1675 & 0 & 0 & 0 & 0 \\ 
		\hline
		 &\multirow{4}{*}{$N=1000$} & CI Length & 0.09881 & 0.1024 & 0.1054 & 4.978 & 0.03402 & 0.4898 & 0.1018 &  0.1028 \\ 
		 & & RMSE & 0.08764 & 0.02848 & 0.02804 & 0.03864 & 0.02594 & 0.1442 & 0.02565 & 0.02599 \\ 
		& & MAE & 0.08488 & 0.02039 & 0.01987 & 0.04186 & 0.01707 & 0.09828 & 0.01714 & 0.0171 \\ 
		& & Failures & 0 & 0 & 0 & 1997 & 0 & 0 & 0 & 0 \\ 
		\hline
	\end{tabular}
\end{sidewaystable}

% latex table generated in R 3.6.3 by xtable 1.8-4 package
% Fri Oct  2 10:30:47 2020
\begin{sidewaystable}[ht]
	\centering
	\caption{\redd{Median length of the 95\% confidence intervals (Length CI), root mean square error of the estimated treatment effect (RMSE), median absolute error (MAE) and number of models failed for Scenario 2: Unmeasured confounder. Crude RD is estimated by a linear regression adjusting only for treatment assignment, Cov adjustment additionally adjusts for baseline covariates, PS covariate denotes the method including the PS in the outcome regression model and Simple DR g-comp and DR using quintiles refer to the doubly robust g-computation methods, respectively}.}
	\label{tab:Scen2}	
	\begin{tabular}{ccccccccccc}
		\hline
		true RD & $N$ &  & Crude RD & \makecell{covariate \\ adjusted} & PS cov & matching & IPTW & \makecell{Simple DR \\g-comp} & \makecell{\red{DR using} \\ \red{quintiles}}  & AIPW \\  
		\hline
		  	\multirow{4}{*}{0} & \multirow{4}{*}{$N=40$}  & CI Length & 0.6118 & 0.7142 & 0.7028 & 3.566 & 0.2116 & 1.15 & 0.7985 & 0.5803 \\ 
		  & & RMSE & 0.3781 & 0.3417 & 0.3416 & 0.363 & 0.3483 & 0.4579 & 0.3475 & 0.347 \\ 
		  & & MAE & 0.3485 & 0.297 & 0.3005 & 0.3077 & 0.3031 & 0.3741 & 0.3032 & 0.2975 \\ 
		  & & Failures & 0 & 0 & 0 & 1255 & 0 & 0 & 0 & 47 \\ 
		  \hline
		  & \multirow{4}{*}{$N=100$} & CI Length & 0.3761 & 0.4038 & 0.4083 & 3.732 & 0.117 & 1.056 & 0.4009 & 0.3738 \\ 
		  & & RMSE & 0.3581 & 0.3143 & 0.3144 & 0.3193 & 0.3147 & 0.4389 & 0.3152 & 0.3146 \\ 
		  & & MAE & 0.3474 & 0.3032 & 0.3033 & 0.3043 & 0.3004 & 0.3504 & 0.3002 & 0.3005 \\ 
		  & & Failures & 0 & 0 & 0 & 1502 & 0 & 0 & 0 & 0 \\ 
		  \hline
		  & \multirow{4}{*}{$N=1000$} & CI Length & 0.1169 & 0.1204 & 0.1233 & 3.768 & 0.03334 & 0.7931 & 0.1197 & 0.1204 \\ 
		  & & RMSE & 0.3471 & 0.2988 & 0.2989 & 0.3083 & 0.2987 & 0.3483 & 0.2983 & 0.2986 \\ 
		  & & MAE & 0.346 & 0.2981 & 0.298 & 0.301 & 0.2978 & 0.2782 & 0.2976 & 0.2979 \\ 
		  & & Failures & 0 & 0 & 0 & 1985 & 0 & 0 & 0 & 0 \\ 
		  \hline
		  \hline
		  	\multirow{4}{*}{0.16} & \multirow{4}{*}{$N=40$}  & CI Length & 0.5639 & 0.6632 & 0.6563 & 3.823 & 0.1863 & 1.15 & 0.7525 & 0.5399 \\ 
		  & & RMSE & 0.3553 & 0.3243 & 0.3247 & 0.3372 & 0.3314 & 0.4361 & 0.3253 & 0.3297 \\ 
		  & & MAE & 0.329 & 0.2889 & 0.2901 & 0.2945 & 0.2891 & 0.41 & 0.2851 & 0.2903 \\ 
		  & & Failures & 0 & 0 & 0 & 1255 & 0 & 0 & 0 & 47 \\
		  \hline 
		  & \multirow{4}{*}{$N=100$} & CI Length & 0.3454 & 0.3761 & 0.3791 & 3.837 & 0.103 & 1.059 & 0.3729 & 0.3485 \\ 
		  & & RMSE & 0.3419 & 0.3056 & 0.3056 & 0.3111 & 0.3046 & 0.4174 & 0.3057 & 0.3046 \\ 
		  & & MAE & 0.3304 & 0.2902 & 0.2904 & 0.2958 & 0.2903 & 0.3981 & 0.2901 & 0.2913 \\ 
		  & & Failures & 0 & 0 & 0 & 1502 & 0 & 0 & 0 & 0 \\ 
		  \hline
		  & \multirow{4}{*}{$N=1000$} & CI Length & 0.1075 & 0.1125 & 0.1149 & 3.85 & 0.02938 & 0.801 & 0.1123 & 0.1127 \\ 
		 & & RMSE & 0.3295 & 0.2888 & 0.2888 & 0.2885 & 0.2875 & 0.3276 & 0.2871 & 0.2874 \\ 
		  & & MAE & 0.3284 & 0.2881 & 0.288 & 0.2889 & 0.2863 & 0.2847 & 0.2866 & 0.2861 \\ 
		  & & Failures & 0 & 0 & 0 & 1985 & 0 & 0 & 0 & 0 \\ 
		  \hline
		  \hline
		  	\multirow{4}{*}{0.4} & \multirow{4}{*}{$N=40$}  & CI Length & 0.4687 & 0.5579 & 0.546 & 3.859 & 0.1567 & 1.15 & 0.5525 & 0.4621 \\ 
		  & & RMSE & 0.2642 & 0.2481 & 0.2486 & 0.2593 & 0.2533 & 0.4077 & 0.2475 & 0.2559 \\ 
		  & & MAE & 0.2429 & 0.2132 & 0.211 & 0.2154 & 0.2145 & 0.4119 & 0.2134 & 0.2149 \\ 
		  & & Failures & 0 & 1 & 0 & 1278 & 0 & 0 & 0 & 51 \\ 
		  \hline
		 & \multirow{4}{*}{$N=100$} & CI Length & 0.2908 & 0.3193 & 0.3185 & 3.815 & 0.0913 & 1.039 & 0.3235 & 0.2982 \\ 
		  & & RMSE & 0.242 & 0.2187 & 0.2186 & 0.2212 & 0.218 & 0.363 & 0.2167 & 0.2178 \\ 
		  & & MAE & 0.2303 & 0.2034 & 0.2036 & 0.2111 & 0.2044 & 0.3615 & 0.2009 & 0.203 \\ 
		  & & Failures & 0 & 0 & 0 & 1513 & 0 & 0 & 0 & 0 \\ 
		  \hline
		 & \multirow{4}{*}{$N=1000$} & CI Length & 0.09057 & 0.09519 & 0.09628 & 3.632 & 0.02705 & 0.7431 & 0.09586 & 0.09596 \\ 
		 & & RMSE & 0.2335 & 0.2077 & 0.2076 & 0.2197 & 0.2046 & 0.2573 & 0.2042 & 0.2046 \\ 
		  & & MAE & 0.2327 & 0.207 & 0.2068 & 0.2274 & 0.2032 & 0.2168 & 0.2032 & 0.2031 \\ 
		  & & Failures & 0 & 0 & 0 & 1990 & 0 & 0 & 0 & 0 \\ 
		 \hline
	\end{tabular}
\end{sidewaystable}

 %latex table generated in R 3.6.3 by xtable 1.8-4 package
% Fri Oct  2 11:41:37 2020
\begin{sidewaystable}[ht]
	\centering
	\caption{\redd{Median length of the 95\% confidence intervals (Length CI), root mean square error of the estimated treatment effect (RMSE), median absolute error (MAE) and number of models failed for Scenario 3: Austin. Crude RD is estimated by a linear regression adjusting only for treatment assignment, Cov adjustment additionally adjusts for baseline covariates, PS covariate denotes the method including the PS in the outcome regression model and Simple DR g-comp and DR using quintiles refer to the doubly robust g-computation methods, respectively}.}
	\label{tab:Scen3}	
	\begin{tabular}{ccccccccccc}
		\hline
		true RD & $N$ &  & Crude RD & \makecell{covariate \\ adjusted} & PS cov & matching & IPTW & \makecell{Simple DR \\g-comp} & \makecell{\red{DR using} \\ \red{quintiles}}  & AIPW \\ 
		\hline
		 	\multirow{4}{*}{0} & \multirow{4}{*}{$N=40$}  & CI Length & 0.5603 & 0.762 & 0.7935 & 1.715 & 0.1818 & 1.125 & 0.925 & 1.126 \\ 
		 & & RMSE & 0.194 & 0.1726 & 0.1945 & 0.2163 & 0.1973 & 0.351 & 0.235 & 0.2144 \\ 
		 & & MAE & 0.15 & 0.1132 & 0.1211 & 0.1429 & 0.1343 & 0.3 & 0.2 & 0.1389 \\ 
		 & & Failures & 0 & 0 & 0 & 982 & 0 & 0 & 0 & 272 \\ 
		 \hline
		 & \multirow{4}{*}{$N=100$} & CI Length & 0.3439 & 0.3961 & 0.4331 & 1.887 & 0.1601 & 0.8337 & 0.4516 & 0.4447 \\ 
		 & & RMSE & 0.1592 & 0.09837 & 0.1008 & 0.1221 & 0.1233 & 0.2284 & 0.1036 & 0.119 \\ 
		 & & MAE & 0.1345 & 0.06787 & 0.06938 & 0.08696 & 0.08036 & 0.1498 & 0.07362 & 0.07635 \\ 
		 & & Failures & 0 & 0 & 0 & 788 & 0 & 0 & 0 & 0 \\ 
		 \hline
		 & \multirow{4}{*}{$N=1000$} & CI Length & 0.1068 & 0.114 & 0.1267 & 1.939 & 0.06029 & 0.2382 & 0.1097 & 0.1366 \\ 
		 & & RMSE & 0.1356 & 0.02942 & 0.02966 & 0.04048 & 0.03514 & 0.06127 & 0.02884 & 0.03374 \\ 
		 & & MAE & 0.133 & 0.01964 & 0.01961 & 0.02799 & 0.02342 & 0.04113 & 0.01954 & 0.02237 \\ 
		 & & Failures & 0 & 0 & 0 & 600 & 0 & 0 & 0 & 0 \\ 
		 \hline
		 \hline
		 	\multirow{4}{*}{0.16} & \multirow{4}{*}{$N=40$}  & CI Length & 0.5831 & 0.8028 & 0.8645 & 2.195 & 0.2533 & 1.125 & 0.9497 & 1.074 \\ 
		 & & RMSE & 0.2098 & 0.1824 & 0.2139 & 0.2327 & 0.2145 & 0.3532 & 0.2429 & 0.2261 \\ 
		 & & MAE & 0.1582 & 0.1217 & 0.131 & 0.16 & 0.1505 & 0.3209 & 0.1789 & 0.1484 \\ 
		 & & Failures & 0 & 0 & 0 & 982 & 0 & 0 & 0 & 272 \\ 
		 \hline
		 & \multirow{4}{*}{$N=100$} & CI Length & 0.3582 & 0.4209 & 0.471 & 2.42 & 0.2227 & 0.8592 & 0.4589 & 0.4449 \\ 
		 & & RMSE & 0.1764 & 0.1033 & 0.1059 & 0.1281 & 0.138 & 0.2297 & 0.1103 & 0.1286 \\ 
		 & & MAE & 0.1533 & 0.06929 & 0.07238 & 0.08593 & 0.08858 & 0.1635 & 0.07497 & 0.08366 \\ 
		 & & Failures & 0 & 0 & 0 & 788 & 0 & 0 & 0 & 0 \\ 
		 \hline
		 & \multirow{4}{*}{$N=1000$} & CI Length & 0.1115 & 0.1214 & 0.1378 & 2.472 & 0.07675 & 0.2576 & 0.1191 & 0.1444 \\ 
		 & & RMSE & 0.1529 & 0.03138 & 0.03129 & 0.0446 & 0.03796 & 0.06673 & 0.03052 & 0.03651 \\ 
		 & & MAE & 0.1505 & 0.02178 & 0.02165 & 0.0316 & 0.02533 & 0.043 & 0.02084 & 0.02467 \\ 
		 & & Failures & 0 & 0 & 0 & 584 & 0 & 0 & 0 & 0 \\ 
		 \hline
		 \hline
		 	\multirow{4}{*}{0.4} & \multirow{4}{*}{$N=40$}  & CI Length & 0.5439 & 0.7943 & 0.867 & 2.695 & 0.2462 & 1.125 & 0.925 & 0.9721 \\ 
		 & & RMSE & 0.1965 & 0.1781 & 0.2048 & 0.2196 & 0.2123 & 0.3738 & 0.2159 & 0.2215 \\ 
		 & & MAE & 0.1514 & 0.1198 & 0.1323 & 0.15 & 0.1578 & 0.3061 & 0.1277 & 0.1423 \\ 
		 & & Failures & 0 & 0 & 0 & 982 & 0 & 0 & 0 & 272 \\ 
		 \hline
		 & \multirow{4}{*}{$N=100$} & CI Length & 0.3342 & 0.4202 & 0.4732 & 2.846 & 0.2415 & 0.9199 & 0.4436 & 0.4277 \\ 
		 & & RMSE & 0.1692 & 0.1036 & 0.106 & 0.1225 & 0.1442 & 0.2462 & 0.1115 & 0.1276 \\ 
		 & & MAE & 0.1484 & 0.06843 & 0.0715 & 0.08 & 0.09595 & 0.1785 & 0.07564 & 0.08568 \\ 
		 & & Failures & 0 & 0 & 0 & 788 & 0 & 0 & 0 & 0 \\ 
		 \hline
		 & \multirow{4}{*}{$N=1000$} & CI Length & 0.1036 & 0.1219 & 0.1387 & 2.863 & 0.09316 & 0.2694 & 0.1262 & 0.1481 \\ 
		 & & RMSE & 0.1486 & 0.03536 & 0.03439 & 0.04698 & 0.04086 & 0.06804 & 0.03299 & 0.03778 \\ 
		 & & MAE & 0.1463 & 0.02417 & 0.02408 & 0.03327 & 0.02782 & 0.04509 & 0.02271 & 0.02512 \\ 
		& & Failures & 0 & 0 & 0 & 584 & 0 & 0 & 0 & 0 \\ 
		\hline
	\end{tabular}
\end{sidewaystable}

Across all scenarios considered here, we note that the matching procedure is the most prone to failure. 
\redd{Even for the large sample sizes, it often fails in creating a matched sample. This is even more pronounced for the situations with unbalanced treatment allocation, see Table~\ref{S1}.}
These results are in line with the findings of \red{Adrillon et al.}~\cite{Andrillon2020}, who stress the need for development of appropriate matching methods in small sample studies. 
Furthermore, we found that using the default caliper, which is 0 in R, leads to extremely biased results with coverage probabilities dropping below 1\% in some situations (results not shown). 

\redd{We note that the mean bias of all methods decreases with growing sample sizes, although the difference is not pronounced.} The largest mean bias is observed for \redd{the crude RD estimation. For Scenario 1 with a true risk difference of 0, however, the AIPW method has the largest bias, see Figure~\ref{fig:bias}. Our simulations also show very good results for simple covariate adjustment with respect to both RMSE and MAE.}

\redd{With respect to coverage, we observe surprisingly poor coverage probabilities for IPTW (Figure~\ref{fig:cov}) and at the same time very short confidence intervals (Tables~\ref{tab:Scen1}--\ref{tab:Scen3}). For the small sample sizes, the coverage of IPTW is even worse than the crude risk difference. The other methods show similar results except for AIPW, which again can not handle Scenario 1 for a risk difference of 0 very well.}

\redd{As expected, the results observed for Scenario 2 show a larger bias and much smaller coverage probabilities than in the other scenarios due to the unobserved confounder. Here, the simple DR g-computation performs best both with respect to bias and coverage probabilities. Coverage probabilities are also very high for matching, but due to the many failures and the extremely wide confidence intervals of the method, these results should be interpreted with caution.}

\begin{figure}[ht]
	\includegraphics[width=\textwidth]{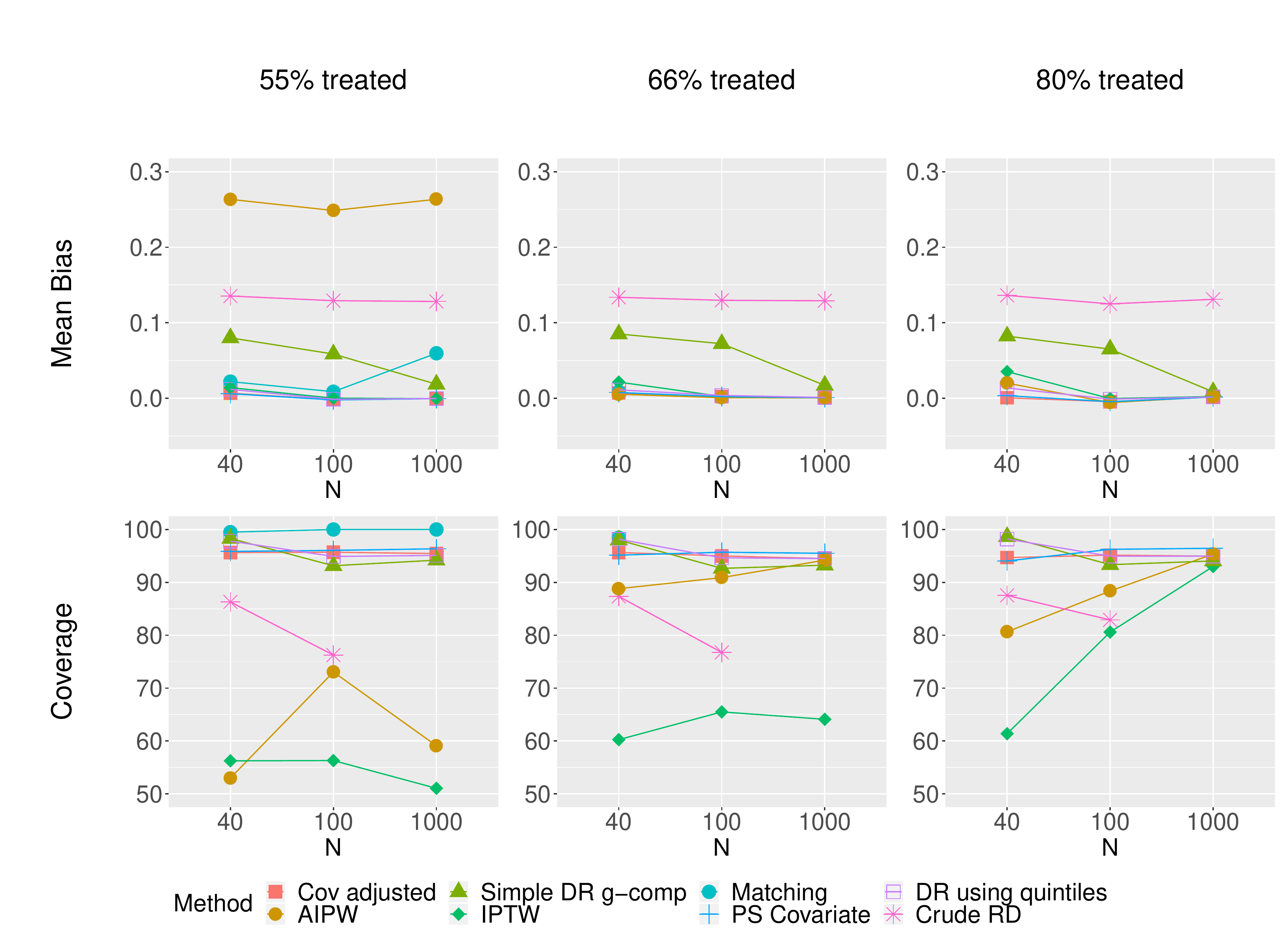}
	\caption{\redd{Mean bias and coverage probabilities for Scenario 1 with a true RD of 0 and different proportions of treated individuals. Note that the coverage is truncated to $\geq 50\%$ implying that the unadjusted method is not displayed for $N=1000$.}}
	\label{fig:Scen1_unb}
\end{figure}

\begin{figure}[ht]
	\centering
	\includegraphics[width=\textwidth]{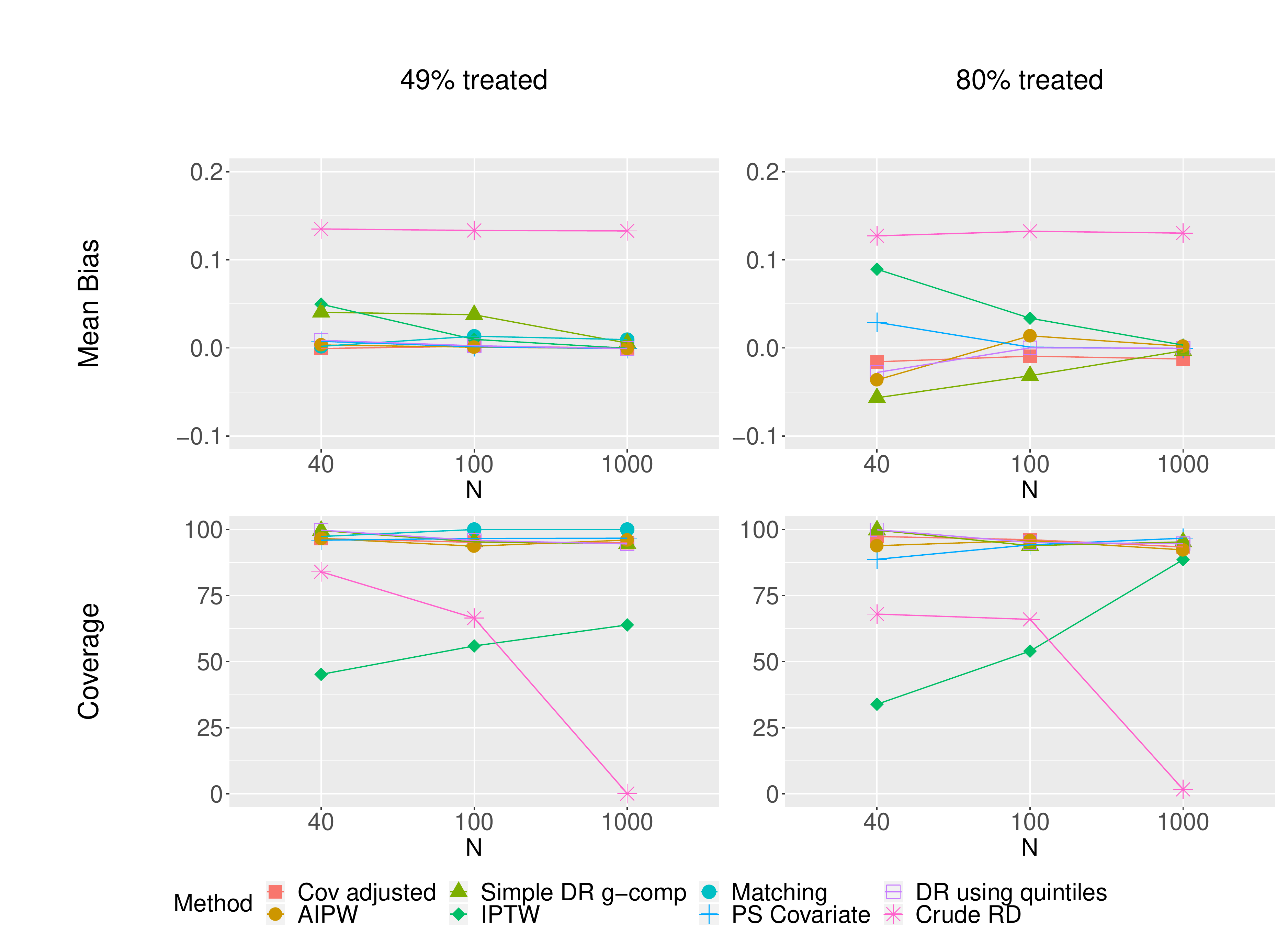}
\caption{\redd{Mean bias and coverage probabilities for Scenario 3 with a true RD of 0 and different proportions of treated individuals.}}
	\label{fig:Scen3_unb}
\end{figure}

Figures~\ref{fig:Scen1_unb} and \ref{fig:Scen3_unb} show the mean bias and coverage probabilities of the methods in Scenario 1 and 3, respectively, where we varied the proportion of individuals who receive treatment. \redd{The results are very similar to the ones observed for the balanced situation.}

\redd{Concerning the doubly robust estimators, we find that the DR using quintiles performs best across Scenarios 1 and 3. Only in Scenario 2, the simple DR g-computation shows the best results. However, more research is needed here to investigate whether this stems from different biases in opposite directions which compensate each other.}

\redd{A few comments on the comparison between OR and RD are in place: As can be seen from the results in the supplemental material, the logistic regression is very unstable for small sample sizes. Thus, we observe a lot more failures than we did for the RD, where only matching and (for small samples) AIPW showed relevant failures. Moreover, the estimation of the OR is sometimes heavily biased, resulting in a large mean bias as opposed to a relatively mild median bias, which in turn also leads to huge confidence intervals.}

% latex table generated in R 3.6.3 by xtable 1.8-4 package
% Fri Oct  2 12:16:21 2020
\begin{sidewaystable}[ht]
	\centering
	\caption{\redd{Median length of the 95\% CIs (Length CI), RMSE, MAE and number of failures for Scenario 1: COVID-19 and Scenario 3: Austin with unbalanced treatment allocation.}}
	\label{S1}	
	\begin{tabular}{ccccccccccc}
		\hline
		& $N$ &  & Crude RD & \makecell{covariate \\ adjusted} & PS cov & matching & IPTW & \makecell{Simple DR \\g-comp} & \makecell{DR using \\ quintiles}  & AIPW \\ 
		\hline
		%\multirow{4}{*}{55\% treated}&		\multirow{4}{*}{$N=40$}& CI Length &  0.6442 & 0.7157 & 0.736 & 4.162 & 0.2569 & 1.15 & 0.8127 & 0.5355 \\ 
		%&	&	RMSE &  0.2105 & 0.169 & 0.1722 & 0.2185 & 0.1847 & 0.39 & 0.1886 & 0.2662 \\ 
		%&	&	MAE &  0.15 & 0.1133 & 0.1142 & 0.1538 & 0.1246 & 0.3695 & 0.1265 & 0.2656 \\ 
		%&	&	failures &  0 & 0 & 0 & 1404 & 0 & 1 & 0 & 0 \\ 
		%	\hline
		%&	\multirow{4}{*}{$N=100$}&	CI Length &  0.397 & 0.4072 & 0.4251 & 4.219 & 0.1463 & 1.001 & 0.4006 & 0.5371 \\ 
		%&	&	RMSE &  0.1621 & 0.09911 & 0.09948 & 0.1206 & 0.103 & 0.3388 & 0.1001 & 0.2497 \\ 
		%&	&	MAE &  0.1319 & 0.06541 & 0.06507 & 0.08696 & 0.06772 & 0.2904 & 0.06678 & 0.2493 \\ 
		%&	&	failures &  0 & 0 & 0 & 1647 & 0 & 0 & 0 & 0 \\ 
		%	\hline
		%&	\multirow{4}{*}{$N=1000$}&	CI Length &  0.1236 & 0.121 & 0.1265 & 4.099 & 0.04152 & 0.5677 & 0.1179 & 0.5354 \\ 
		%&	&	RMSE &  0.1319 & 0.0304 & 0.03046 & 0.06247 & 0.03058 & 0.1609 & 0.03002 & 0.2636 \\ 
		%&	&	MAE &  0.1278 & 0.02041 & 0.02068 & 0.0596 & 0.02063 & 0.1054 & 0.02028 & 0.2632 \\ 
		%&	&	failures &  0 & 0 & 0 & 1998 & 0 & 0 & 0 & 0 \\ 
		%	\hline
		&	\multirow{4}{*}{$N=40$}&	CI Length &  0.6684 & 0.752 & 0.7806 & 4.074 & 0.3096 & 1.15 & 0.8421 & 0.6278 \\ 
		Scenario 1 &	&	RMSE &  0.2136 & 0.176 & 0.1795 & 0.2295 & 0.1987 & 0.3823 & 0.1961 & 0.2544 \\ 
		66\% treated &	&	MAE &  0.1535 & 0.117 & 0.1182 & 0.125 & 0.1331 & 0.3484 & 0.1321 & 0.1307 \\ 
		&	&	failures &  0 & 1 & 0 & 1936 & 0 & 1 & 0 & 60 \\ 
		\hline
		&	\multirow{4}{*}{$N=100$}&	CI Length &  0.4158 & 0.4264 & 0.4463 & $\cdot$ & 0.1949 & 0.9479 & 0.4172 & 0.3925 \\ 
		&	&	RMSE &  0.1674 & 0.1054 & 0.1061 & $\cdot$ & 0.114 & 0.3129 & 0.1067 & 0.1134 \\ 
		&	&	MAE &  0.1298 & 0.06967 & 0.0703 & $\cdot$ & 0.07415 & 0.2511 & 0.07192 & 0.07421 \\ 
		&	&	failures &  0 & 0 & 0 & 2000 & 0 & 0 & 0 & 0 \\ 
		\hline
		&	\multirow{4}{*}{$N=1000$}&	CI Length &  0.1298 & 0.1258 & 0.1323 & $\cdot$ & 0.05847 & 0.4306 & 0.1228 & 0.1277 \\ 
		&	&	RMSE &  0.1333 & 0.0322 & 0.03225 & $\cdot$ & 0.03336 & 0.1177 & 0.03158 & 0.03311 \\ 
		&	&	MAE &  0.129 & 0.02163 & 0.02188 & $\cdot$ & 0.02199 & 0.07594 & 0.0214 & 0.02202 \\ 
		&	&	failures &  0 & 0 & 0 & 2000 & 0 & 0 & 0 & 0 \\ 
		\hline
		&	\multirow{4}{*}{$N=40$}&	CI Length &  0.8139 & 0.8853 & 0.9398 & $\cdot$ & 0.4916 & 1.171 & 0.9248 & 0.8298 \\ 
		Scenario 1 &	&	RMSE &  0.2454 & 0.2088 & 0.2178 & $\cdot$ & 0.2577 & 0.3901 & 0.2345 & 0.8889 \\ 
		80\% treated &	&	MAE &  0.1714 & 0.1376 & 0.1448 & $\cdot$ & 0.1853 & 0.3552 & 0.16 & 0.1929 \\ 
		&	&	failures &  1 & 2 & 1 & 2000 & 1 & 0 & 0 & 605 \\ 
		\hline
		&	\multirow{4}{*}{$N=100$}&	CI Length &  0.4956 & 0.4942 & 0.526 & $\cdot$ & 0.4647 & 0.9358 & 0.4874 & 0.4724 \\ 
		&	&	RMSE &  0.1761 & 0.1218 & 0.1239 & $\cdot$ & 0.1516 & 0.2775 & 0.1245 & 0.1569 \\ 
		&	&	MAE &  0.1294 & 0.08333 & 0.08371 & $\cdot$ & 0.1018 & 0.2026 & 0.08554 & 0.09977 \\ 
		&	&	failures &  0 & 0 & 0 & 2000 & 0 & 0 & 0 & 7 \\ 
		\hline
		&	\multirow{4}{*}{$N=1000$}&	CI Length &  0.1531 & 0.1458 & 0.1546 & $\cdot$ & 0.1476 & 0.3277 & 0.143 & 0.1599 \\ 
		&	&	RMSE &  0.1365 & 0.03668 & 0.03673 & $\cdot$ & 0.0409 & 0.08494 & 0.03639 & 0.04032 \\ 
		&	&	MAE &  0.1319 & 0.02492 & 0.02469 & $\cdot$ & 0.02813 & 0.05294 & 0.02466 & 0.02801 \\ 
		&	&	failures &  0 & 0 & 0 & 2000 & 0 & 0 & 0 & 0 \\ 
		\hline
		\hline
		&	\multirow{4}{*}{$N=40$}&	CI Length & 0.6086 & 0.8389 & 0.8496 & $\cdot$ & 0.1076 & 1.15 & 0.975 & 1.829 \\ 
		Scenario 3 &	&	RMSE & 0.1992 & 0.1929 & 0.2359 & $\cdot$ & 0.2228 & 0.4026 & 0.2751 & 0.4225 \\ 
		80\% treated &	&	MAE & 0.1562 & 0.1223 & 0.1352 & $\cdot$ & 0.1785 & 0.35 & 0.225 & 0.2219 \\ 
		&	&	failures & 3 & 3 & 3 & 2000 & 3 & 0 & 0 & 1690 \\ 
		\hline
		&	\multirow{4}{*}{$N=100$} &	CI Length & 0.3744 & 0.4348 & 0.4804 & $\cdot$ & 0.2419 & 0.8949 & 0.5733 & 0.9399 \\ 
		&	&	RMSE & 0.1636 & 0.1029 & 0.1092 & $\cdot$ & 0.2022 & 0.2403 & 0.1412 & 0.2176 \\ 
		&	&	MAE & 0.1382 & 0.06706 & 0.07216 & $\cdot$ & 0.1492 & 0.1709 & 0.09435 & 0.1439 \\ 
		&	&	failures & 0 & 0 & 0 & 2000 & 0 & 0 & 0 & 45 \\ 
		\hline
		&	\multirow{4}{*}{$N=1000$} &	CI Length & 0.1166 & 0.1226 & 0.136 & $\cdot$ & 0.439 & 0.2122 & 0.146 & 0.2318 \\ 
		&	&	RMSE & 0.1338 & 0.03366 & 0.03193 & $\cdot$ & 0.07436 & 0.05153 & 0.03761 & 0.06406 \\ 
		&	&	MAE & 0.1304 & 0.02278 & 0.0222 & $\cdot$ & 0.04846 & 0.03508 & 0.02573 & 0.04318 \\ 
		&	&	failures & 0 & 0 & 0 & 2000 & 0 & 0 & 0 & 0 \\ 
		\hline
	\end{tabular}
\end{sidewaystable}

\subsection{Recommendations for small-scale studies}
Based on the simulation results, we deduct the following recommendations for applications in clinical studies:
\begin{enumerate}
	\item \redd{Causal inference methods can correct for the \emph{non-randomized} nature of a study. However, they cannot deal with other issues such as flaws in the study design, data quality etc. This is nicely demonstrated by the simulation results observed for Scenario 2: The presence of an unmeasured confounder renders the methods practically useless. Thus, to speak with Rubin's words, the most important recommendation is: `` For objective causal inference, design trumps analysis''\citep{rubin2008objective}.}
	
	\item \redd{In small sample settings, the risk difference provides a more stable effect measure than the odds ratio, which is due to the limitations of the logistic regression in small samples. Thus, the risk difference is the preferred effect measure in small samples. This recommendation does not only apply to the causal inference methods but also to covariate adjustment.}

	\item For small total sample sizes ($N=40$), \redd{the best performance was observed for covariate adjustment, PS covariate and the DR using quintiles, i.~e.~a doubly robust g-computation.}
	\item \redd{IPTW performed well with respect to bias, RMSE and MAE, but due to its extremely low coverage probability it can not be recommended.}
	%\item With an unobserved confounder, the simple DR g-computation provided the best results with respect to coverage and bias. \redd{As mentioned above, this result warrants further investigation.}
	\item Since one can always only investigate a limited number of simulated settings, we recommend conducting simulations for a given example at hand. In order to facilitate this, the R code used for the simulations in this paper is available from Github (\url{https://github.com/smn74/CIM_COVID-19}).
\end{enumerate}

\section{Discussion} \label{sec:dis}
%(max. 1500 words)

In ongoing pandemics there is an urgent unmet medical need to develop vaccines, diagnostics and treatments in a very timely fashion. Despite the time pressure, however, the standards of evidence should not unduly be lowered \citep{Bauchner2020,ferreira2020decline,muetze2020,Rome2020}. Using a small-scale non-randomized study in COVID-19 \cite{gautret2020hydroxychloroquine} as a motivating example, we \redd{discuss} how robust analyses can be conducted by use of appropriate causal inference methods.
%We speculate that application of doubly-robust g-computation in the motivating study might have cast doubt on the efficacy of hydroxychloroquine and consequently, in the context of other criticism voiced, might have dampened the early enthusiasm regarding the use of this drug in COVID-19. Ultimately this might have saved some resources that now were wasted on clinical trials investigating hydroxychloroquine in patients suffering from COVID-19. More importantly, this might have prevented a shortage of hydroxychloroquine in the treatment of rheumatological disorders, which some tried to counter by revised treatment schedules to maintain the standard of care \citep{Scheetz2020}.

The conventional method to correct for baseline differences between groups is adjusting for all relevant patient characteristics in the outcome regression model. This is, however, not favorable for different reasons. As \red{Rosenbaum and Rubin}~\cite{Rosenbaum1983} point out, covariate adjustment works poorly in cases where e.~g.~the variance of a covariate is unequal in the treatment and the control group. A commonly applied alternative in observational studies are propensity score methods. Since these methods were derived from a formal model for causal inference, their use allows for well-defined causal questions \citep{goetghebeur2020,McCaffrey2013}.  Moreover, propensity score methods also work as a dimension reduction tool by combining multiple covariates into a single score \citep{McCaffrey2013,Rosenbaum1984}. This is especially important in situations with a large number of covariates compared to the number of subjects under study. %In our data example, adjusting for all baseline covariates in the outcome model was not \red{possible without applying a correction like Firth's penalization} due to the small number of subjects in the two groups. For comparison, we therefore included the special case of a simple logistic regression model with treatment assignment as the only covariate.

Different approaches have been suggested for the PS modeling strategy. Originally, nonparsimonious models including all potential confounders have been recommended for the propensity score \cite{rubin1996matching}. This approach, however, may not be feasible in small samples. Thus, it has been recommended \citep{Andrillon2020,brookhart2006variable,pirracchio2012evaluation} to use some kind of variable selection procedure in this case, but clear recommendations are lacking \citep{goetghebeur2020}. Importantly, the choice of the variables should not be based on some goodness-of-fit measure \citep{brookhart2006variable,Weitzen_2004} but rather on the relationship of the variables with both the outcome and the exposure \cite{pirracchio2012evaluation}. Since our data example only contained four potential confounders, we have included them all in the PS model and have not investigated methods of variable selection here.

In line with \cite{kang2007demystifying} our simulation studies showed that different DR estimators led to different results in the scenarios considered. More thorough investigations on this topic, especially concerning the type of DR adjustment, will be part of future research.

Some comments on the estimands obtained by the different methods are in place: 
\redd{First,} our aim was to estimate the average causal effect in our study population. 
\red{When there is a lack of overlap between the propensity score distributions of the two groups, a problem quite common for small samples, IPTW may become unstable due to extremely large weights. Mao et al.~\cite{mao2019propensity} recently proposed  modified weights, which result in a different estimand that deviates from the average treatment effect. However, this approach is appropriate in treatment effect discovery, which is often the main motivation of small observational studies. Similarly,} propensity score matching creates a population where treated individuals, who cannot be matched to any control patients, are excluded. Thus, the effect estimate obtained here corresponds to a subset of the population, which is hard to describe. Since the matched population is not very well characterized, it is difficult to generalize results obtained there to the general population \citep{Hernan2020}. \red{Moreover, PS matching is also criticized for the fact that a large number of irrelevant covariates might lead to matched pairs which actually differ in relevant covariates \citep{dieng2019interpretable,wang2017flame}. Solutions to this problem use machine learning techniques to first determine the relevant covariates and then match exactly on these \citep{dieng2019interpretable,wang2017flame}. Since they require large training and matching sets, however, they can not be applied to small samples.}
\red{Furthermore, we did not account for the fact that the propensity score used for matching is itself estimated, see \cite{Abadie2016} for a thorough discussion of this topic. Moreover, it should be noted that classical bootstrap approaches such as the nonparametric bootstrap we used in the g-computation are not applicable to matching estimators \citep{Abadie2006,Abadie2008}.}
It is also worth noting that among the methods we discussed here, only IPT weighting and g-computation can be generalized to more complex situations involving time-varying treatments \citep{Hernan2020}.
\redd{Finally, it has to be noted that when estimating the odds ratio instead of the risk difference, \emph{marginal} and \emph{conditional} treatment effects differ due to non-collapsibility of the odds ratio~\citep{Austin_etal_2007,Mao2020}. It should be noted that covariate adjusted logistic regression, PS covariate adjusted logistic regression, and conditional logistic regression in the matched sample all estimate the conditional OR and not the marginal OR in this case. Thus, care has to be taken when comparing the results of the different PS methods.
}

Motivated by the study conducted by \red{Gautret et al.}~\cite{gautret2020hydroxychloroquine} we investigated the properties of a range of causal inference methods in small samples. As expected this posed additional challenges to the various approaches. Interestingly, it turned out that the default settings in software implementations are often more suitable for large sample sizes and need to be adjusted for applications in small-scale studies. For example, we found that the matching procedure in R using the default calipers of 0 resulted in extremely biased results in our small sample simulations. SAS software, in contrast, uses a default caliper width of 0.25. \red{The issue of choosing the right caliper width has recently been investigated by Wang~\cite{wang2020}, who recommended to take both matching and population bias into account.}

\redd{We did not discuss the (causal) assumptions underlying the different estimation methods proposed in this paper. The recently published tutorial by Goetghebeur et al.~\cite{goetghebeur2020} provides a general overview of these assumptions and how the methods discussed here invoke them. However, they also caution against the possible complications an applied statistician might face when conducting a causal analysis. In particular, it is not sufficient to focus on the non-randomized nature of a study and ignore, e.~g.~design issues, measurement error or study discontinuation, to name a few. This is exactly the case with our motivating data example. Focusing only on the non-randomized nature of the study by Gautret et al.~\cite{gautret2020hydroxychloroquine}, our reanalysis of the study (results not shown, code can be found on Github) was disappointing in that the conclusions based on various considered  approaches did no differ from those reported by Gautret et al.~\cite{gautret2020hydroxychloroquine} that we sought to correct and that disagree with recent large-scale trials \citep{Sattui2020}. Thus, our results demonstrate that while the causal inference methods can provide adjustment for baseline covariates, even a correctly applied causal inference method can not compensate for design issues of the underlying study such as
% However, the study suffers also from other weaknesses including
 the small sample size, open label treatment and study discontinuations \citep{rubin2008objective}. }
 %For instance, we did not address the problems in the interpretation caused by study discontinuations here, but used the last-observation-carried-forward approach as Gautret et al.~\cite{gautret2020hydroxychloroquine} although this approach has gone out of fashion due to its limitations, in particular in underestimating the standard errors, see e.~g.~\cite{molenberghs2004analyzing} and the references cited therein. 

 Our study has several limitations. \redd{First,}
 while our simulation scenarios are carefully chosen to reflect different situations, we could only consider a limited number of settings. Thus, there is no guarantee that our results can be generalized to different situations.
 	We therefore make our R code available, which can be used to explore specific scenarios.
 \redd{Second}, as is known from the literature, we observed that the logistic regression model often failed for the small sample sizes, especially in combination with matching. To investigate whether PS matching can be improved in small sample sizes by using a penalization method shall be part of future research. 
 Finally, our paper only studied a binary outcome and did not consider other commonly used outcomes in clinical studies such as time-to-event data or continuous outcomes. Especially in the context of time-to-event outcomes and longitudinal data, where time-varying treatments additionally complicate estimation, doubly robust g-computation such as TMLE \citep{van2010collaborative,stitelman2012general} is recommended due to its good statistical properties. \redd{Similar to the issues discussed in this paper, the selection of a suitable endpoint requires some care and it is not always appropriate to use the most common approach, see \cite{mccaw2020selecting} for a recent discussion in the context of time-to-event endpoints in COVID-19.}

Besides the design of efficient trials to develop treatments for COVID-19 \citep{Stallard2020}, one concern to trialists these days is the threat posed by the SARS-CoV-2 pandemic to clinical trials in non-COVID-19 indications \citep{Anker2020, Kunz2020}. SARS-CoV-2 infections of patients in these trials, or merely the increased risk thereof, might lead to post-randomization events (or intercurrent events in the language of the ICH E9 addendum \citep{ICH}) such as treatment or study discontinuations as well as adverse events that ultimately invalidate an analysis relying on randomization. In such situations, the causal inference approach discussed here might provide a suitable alternative analysis strategy either as primary or sensitivity analysis.

\section*{Acknowledgement} 

Support by the DFG (grant FR 4121/2-1) is gratefully acknowledged.

 \section*{Conflict of interest}

The authors declare that they have no conflict of interest.

%\section*{Appendix}

\bibliographystyle{plainurl}
\bibliography{Lit-Covid}

\begin{thebibliography}{10}

\bibitem{aalen}
Odd~O Aalen, Richard~J Cook, and Kjetil R{\o}ysland.
\newblock {Does Cox analysis of a randomized survival study yield a causal
  treatment effect?}
\newblock {\em Lifetime Data Analysis}, 21(4):579--593, 2015.

\bibitem{Abadie2006}
Alberto Abadie and Guido~W Imbens.
\newblock Large sample properties of matching estimators for average treatment
  effects.
\newblock {\em Econometrica}, 74(1):235--267, 2006.

\bibitem{Abadie2008}
Alberto Abadie and Guido~W Imbens.
\newblock On the failure of the bootstrap for matching estimators.
\newblock {\em Econometrica}, 76(6):1537--1557, 2008.

\bibitem{Abadie2016}
Alberto Abadie and Guido~W Imbens.
\newblock Matching on the estimated propensity score.
\newblock {\em Econometrica}, 84(2):781--807, 2016.

\bibitem{Alexander2020}
Paul~Elias Alexander, Victoria~Borg Debono, Manoj~J. Mammen, Alfonso Iorio,
  Komal Aryala, Dianna Deng, Eva Brocard, and Waleed Alhazzani.
\newblock {COVID-19 coronavirus research has overall low methodological quality
  thus far: case in point for chloroquine/hydroxychloroquine}.
\newblock {\em Journal of Clinical Epidemiology}, 123:120--126, 2020.

\bibitem{Althauser1970}
Robert~P Althauser and Donald Rubin.
\newblock The computerized construction of a matched sample.
\newblock {\em American Journal of Sociology}, 76(2):325--346, 1970.

\bibitem{altman1991statistics}
Douglas~G Altman.
\newblock {Statistics in Medical Journals: Developments in the 1980s}.
\newblock {\em Statistics in Medicine}, 10(12):1897--1913, 1991.

\bibitem{altman2000we}
Douglas~G Altman and Patrick Royston.
\newblock What do we mean by validating a prognostic model?
\newblock {\em Statistics in Medicine}, 19(4):453--473, 2000.

\bibitem{Andrillon2020}
Anais Andrillon, Romain Pirracchio, and Sylvie Chevret.
\newblock Performance of propensity score matching to estimate causal effects
  in small samples.
\newblock {\em Statistical Methods in Medical Research}, 29(3):644--658, 2020.

\bibitem{Anker2020}
Stefan~D Anker, Javed Butler, Muhammad~Shahzeb Khan, William~T Abraham, Johann
  Bauersachs, Edimar Bocchi, Biykem Bozkurt, Eugene Braunwald, Vijay~K Chopra,
  John~G Cleland, Justin Ezekowitz, Gerasimos Filippatos, Tim Friede, Adrian~F
  Hernandez, Carolyn S~P Lam, JoAnn Lindenfeld, John J~V McMurray, Mandeep
  Mehra, Marco Metra, Milton Packer, Burkert Pieske, Stuart~J Pocock, Piotr
  Ponikowski, Giuseppe M~C Rosano, John~R Teerlink, Hiroyuki Tsutsui, Dirk~J
  Van~Veldhuisen, Subodh Verma, Adriaan~A Voors, Janet Wittes, Faiez Zannad,
  Jian Zhang, Petar Seferovic, and Andrew J~S Coats.
\newblock {Conducting clinical trials in heart failure during (and after) the
  COVID-19 pandemic: an Expert Consensus Position Paper from the Heart Failure
  Association (HFA) of the European Society of Cardiology (ESC)}.
\newblock {\em European Heart Journal}, 41(22):2109--2117, 2020.
\newblock URL: \url{https://doi.org/10.1093/eurheartj/ehaa461}, \href
  {http://arxiv.org/abs/https://academic.oup.com/eurheartj/article-pdf/41/22/2109/33368354/ehaa461.pdf}
  {\path{arXiv:https://academic.oup.com/eurheartj/article-pdf/41/22/2109/33368354/ehaa461.pdf}},
  \href {http://dx.doi.org/10.1093/eurheartj/ehaa461}
  {\path{doi:10.1093/eurheartj/ehaa461}}.

\bibitem{Austin_2007}
Peter~C. Austin.
\newblock The performance of different propensity score methods for estimating
  marginal odds ratios.
\newblock {\em Statistics in Medicine}, 26(16):3078--3094, 2007.
\newblock URL: \url{https://doi.org/10.1002%2Fsim.2781}, \href
  {http://dx.doi.org/10.1002/sim.2781} {\path{doi:10.1002/sim.2781}}.

\bibitem{Austin2010}
Peter~C Austin.
\newblock The performance of different propensity-score methods for estimating
  differences in proportions (risk differences or absolute risk reductions) in
  observational studies.
\newblock {\em Statistics in Medicine}, 29(20):2137--2148, 2010.

\bibitem{Austin2011}
Peter~C Austin.
\newblock An introduction to propensity score methods for reducing the effects
  of confounding in observational studies.
\newblock {\em Multivariate Behavioral Research}, 46(3):399--424, 2011.

\bibitem{austin2014comparison}
Peter~C Austin.
\newblock A comparison of 12 algorithms for matching on the propensity score.
\newblock {\em Statistics in Medicine}, 33(6):1057--1069, 2014.

\bibitem{Austin_etal_2007}
Peter~C. Austin, Paul Grootendorst, Sharon-Lise~T. Normand, and Geoffrey~M.
  Anderson.
\newblock Conditioning on the propensity score can result in biased estimation
  of common measures of treatment effect: a monte carlo study.
\newblock {\em Statistics in Medicine}, 26(4):754--768, 2007.
\newblock URL: \url{https://doi.org/10.1002%2Fsim.2618}, \href
  {http://dx.doi.org/10.1002/sim.2618} {\path{doi:10.1002/sim.2618}}.

\bibitem{bang2005doubly}
Heejung Bang and James~M Robins.
\newblock Doubly robust estimation in missing data and causal inference models.
\newblock {\em Biometrics}, 61(4):962--973, 2005.

\bibitem{Bauchner2020}
Howard Bauchner and Phil~B. Fontanarosa.
\newblock {Randomized Clinical Trials and COVID-19: Managing Expectations}.
\newblock {\em JAMA}, 323(22):2262--2263, 06 2020.
\newblock URL: \url{https://doi.org/10.1001/jama.2020.8115}, \href
  {http://arxiv.org/abs/https://jamanetwork.com/journals/jama/articlepdf/2765696/jama\_bauchner\_2020\_ed\_200043.pdf}
  {\path{arXiv:https://jamanetwork.com/journals/jama/articlepdf/2765696/jama\_bauchner\_2020\_ed\_200043.pdf}},
  \href {http://dx.doi.org/10.1001/jama.2020.8115}
  {\path{doi:10.1001/jama.2020.8115}}.

\bibitem{Beyersmann2013}
Jan Beyersmann, Susanna Di~Termini, and Markus Pauly.
\newblock Weak convergence of the wild bootstrap for the {A}alen--{J}ohansen
  estimator of the cumulative incidence function of a competing risk.
\newblock {\em Scandinavian Journal of Statistics}, 40(3):387--402, 2013.

\bibitem{bland2000odds}
J~Martin Bland and Douglas~G Altman.
\newblock The odds ratio.
\newblock {\em BMJ}, 320(7247):1468, 2000.

\bibitem{bluhmki2018}
Tobias Bluhmki, Claudia Schmoor, Dennis Dobler, Markus Pauly, J\"urgen Finke,
  Martin Schumacher, and Jan Beyersmann.
\newblock A wild bootstrap approach for the aalen-johansen estimator.
\newblock {\em Biometrics}, 2018.
\newblock To appear.

\bibitem{Boulware2020}
David~R. Boulware, Matthew~F. Pullen, Ananta~S. Bangdiwala, Katelyn~A. Pastick,
  Sarah~M. Lofgren, Elizabeth~C. Okafor, Caleb~P. Skipper, Alanna~A. Nascene,
  Melanie~R. Nicol, Mahsa Abassi, Nicole~W. Engen, Matthew~P. Cheng, Derek
  LaBar, Sylvain~A. Lother, Lauren~J. MacKenzie, Glen Drobot, Nicole Marten,
  Ryan Zarychanski, Lauren~E. Kelly, Ilan~S. Schwartz, Emily~G. McDonald, Radha
  Rajasingham, Todd~C. Lee, and Kathy~H. Hullsiek.
\newblock A randomized trial of hydroxychloroquine as postexposure prophylaxis
  for covid-19.
\newblock {\em New England Journal of Medicine}, 2020.
\newblock URL: \url{https://doi.org/10.1056/NEJMoa2016638}, \href
  {http://arxiv.org/abs/https://doi.org/10.1056/NEJMoa2016638}
  {\path{arXiv:https://doi.org/10.1056/NEJMoa2016638}}, \href
  {http://dx.doi.org/10.1056/NEJMoa2016638} {\path{doi:10.1056/NEJMoa2016638}}.

\bibitem{brookhart2006variable}
M~Alan Brookhart, Sebastian Schneeweiss, Kenneth~J Rothman, Robert~J Glynn,
  Jerry Avorn, and Til St{\"u}rmer.
\newblock Variable selection for propensity score models.
\newblock {\em American Journal of Epidemiology}, 163(12):1149--1156, 2006.

\bibitem{cavalcanti2020}
Alexandre~B. Cavalcanti, Fernando~G. Zampieri, Regis~G. Rosa, Luciano~C.P.
  Azevedo, Viviane~C. Veiga, Alvaro Avezum, Lucas~P. Damiani, Aline Marcadenti,
  Leticia Kawano-Dourado, Thiago Lisboa, Debora L.~M. Junqueira, Pedro~G.M.
  de~Barros~e Silva, Lucas Tramujas, Erlon~O. Abreu-Silva, Ligia~N. Laranjeira,
  Aline~T. Soares, Leandro~S. Echenique, Adriano~J. Pereira, Flavio~G.R.
  Freitas, Otavio~C.E. Gebara, Vicente~C.S. Dantas, Remo~H.M. Furtado,
  Eveline~P. Milan, Nicole~A. Golin, Fabio~F. Cardoso, Israel~S. Maia,
  Conrado~R. Hoffmann~Filho, Adrian~P.M. Kormann, Roberto~B. Amazonas,
  Monalisa~F. Bocchi~de Oliveira, Ary Serpa-Neto, Maicon Falavigna, Renato~D.
  Lopes, Flavia~R. Machado, and Otavio Berwanger.
\newblock Hydroxychloroquine with or without azithromycin in mild-to-moderate
  covid-19.
\newblock {\em New England Journal of Medicine}, 2020.
\newblock URL: \url{https://doi.org/10.1056/NEJMoa2019014}, \href
  {http://arxiv.org/abs/https://doi.org/10.1056/NEJMoa2019014}
  {\path{arXiv:https://doi.org/10.1056/NEJMoa2019014}}, \href
  {http://dx.doi.org/10.1056/NEJMoa2019014} {\path{doi:10.1056/NEJMoa2019014}}.

\bibitem{cheung2007}
Yin~Bun Cheung.
\newblock A modified least-squares regression approach to the estimation of
  risk difference.
\newblock {\em American Journal of Epidemiology}, 166(11):1337--1344, 2007.

\bibitem{cook2002advanced}
Thomas~D Cook.
\newblock {Advanced statistics: up with odds ratios! A case for odds ratios
  when outcomes are common}.
\newblock {\em Academic Emergency Medicine}, 9(12):1430--1434, 2002.

\bibitem{Cortegiani2020}
Andrea Cortegiani, Giulia Ingoglia, Mariachiara Ippolito, Antonino Giarratano,
  and Sharon Einav.
\newblock {A systematic review on the efficacy and safety of chloroquine for
  the treatment of COVID-19}.
\newblock {\em Journal of Critical Care}, 57:279--283, 2020.

\bibitem{cummings2009relative}
Peter Cummings.
\newblock The relative merits of risk ratios and odds ratios.
\newblock {\em Archives of pediatrics \& adolescent medicine}, 163(5):438--445,
  2009.

\bibitem{dieng2019interpretable}
Awa Dieng, Yameng Liu, Sudeepa Roy, Cynthia Rudin, and Alexander Volfovsky.
\newblock Interpretable almost-exact matching for causal inference.
\newblock {\em Proceedings of Machine Learning Research}, 89:2445, 2019.

\bibitem{efron1986bootstrap}
Bradley Efron and Robert Tibshirani.
\newblock Bootstrap methods for standard errors, confidence intervals, and
  other measures of statistical accuracy.
\newblock {\em Statistical Science}, pages 54--75, 1986.

\bibitem{falagas2009well}
Matthew~E Falagas, Gregory~C Makris, Drosos~E Karageorgopoulos, Maria Batsiou,
  and Vangelis~G Alexiou.
\newblock How well do clinical researchers understand risk estimates?
\newblock {\em Epidemiology}, 20(6):930--931, 2009.

\bibitem{ferreira2020decline}
Jo{\~a}o~Pedro Ferreira, Murray Epstein, and Faiez Zannad.
\newblock {The Decline of the Experimental Paradigm During the COVID-19
  Pandemic: A Template for the Future}.
\newblock {\em The American Journal of Medicine}, 2020.
\newblock \href {http://dx.doi.org/10.1016/j.amjmed.2020.08.021}
  {\path{doi:10.1016/j.amjmed.2020.08.021}}.

\bibitem{Friedrich2017wild}
Sarah Friedrich, Frank Konietschke, and Markus Pauly.
\newblock A wild bootstrap approach for nonparametric repeated measurements.
\newblock {\em Computational Statistics \& Data Analysis}, 113:38--52, 2017.

\bibitem{friedrichMATS}
Sarah Friedrich and Markus Pauly.
\newblock {MATS}: Inference for potentially singular and heteroscedastic
  {MANOVA}.
\newblock {\em Journal of Multivariate Analysis}, 165:166--179, 2018.

\bibitem{funck2020}
Christian Funck-Brentano, Lee~S Nguyen, and Joe-Elie Salem.
\newblock {Retraction and republication: cardiac toxicity of hydroxychloroquine
  in COVID-19}.
\newblock {\em The Lancet}, 2020.
\newblock \href {http://dx.doi.org/10.1016/S0140-6736(20)31528-2}
  {\path{doi:10.1016/S0140-6736(20)31528-2}}.

\bibitem{Gail1984}
Mitchell~H Gail, S~Wieand, and Steven Piantadosi.
\newblock Biased estimates of treatment effect in randomized experiments with
  nonlinear regressions and omitted covariates.
\newblock {\em Biometrika}, 71(3):431--444, 1984.

\bibitem{gautret2020hydroxychloroquine}
Philippe Gautret, Jean-Christophe Lagier, Philippe Parola, Line Meddeb, Morgane
  Mailhe, Barbara Doudier, Johan Courjon, Val{\'e}rie Giordanengo, Vera~Esteves
  Vieira, Herv{\'e}~Tissot Dupont, et~al.
\newblock {Hydroxychloroquine and azithromycin as a treatment of COVID-19:
  results of an open-label non-randomized clinical trial}.
\newblock {\em International Journal of Antimicrobial Agents}, 56(1), 2020.
\newblock \href
  {http://dx.doi.org/https://doi.org/10.1016/j.ijantimicag.2020.105949}
  {\path{doi:https://doi.org/10.1016/j.ijantimicag.2020.105949}}.

\bibitem{goetghebeur2020}
Els Goetghebeur, Saskia le~Cessie, Bianca De~Stavola, Erica~EM Moodie, Ingeborg
  Waernbaum, and "on behalf of" the topic group Causal Inference (TG7) of~the
  STRATOS~initiative.
\newblock Formulating causal questions and principled statistical answers.
\newblock {\em Statistics in Medicine}, pages 1--27, 2020.
\newblock URL: \url{https://onlinelibrary.wiley.com/doi/abs/10.1002/sim.8741},
  \href
  {http://arxiv.org/abs/https://onlinelibrary.wiley.com/doi/pdf/10.1002/sim.8741}
  {\path{arXiv:https://onlinelibrary.wiley.com/doi/pdf/10.1002/sim.8741}},
  \href {http://dx.doi.org/10.1002/sim.8741} {\path{doi:10.1002/sim.8741}}.

\bibitem{Hernan2020}
Miguel Hernan and James Robins.
\newblock {\em {Causal Inference: What If}}.
\newblock {Chapman \& Hall/CRC}, Boca Raton, 2020.

\bibitem{Horby2020}
Peter Horby, Marion Mafham, Louise Linsell, Jennifer~L Bell, Natalie Staplin,
  Jonathan~R Emberson, Martin Wiselka, Andrew Ustianowski, Einas Elmahi,
  Benjamin Prudon, Anthony Whitehouse, Timothy Felton, John Williams, Jakki
  Faccenda, Jonathan Underwood, J~Kenneth Baillie, Lucy Chappell, Saul~N Faust,
  Thomas Jaki, Katie Jeffery, Wei~Shen Lim, Alan Montgomery, Kathryn Rowan,
  Joel Tarning, James~A Watson, Nicholas~J White, Edmund Juszczak, Richard
  Haynes, and Martin~J Landray.
\newblock {Effect of Hydroxychloroquine in Hospitalized Patients with COVID-19:
  Preliminary results from a multi-centre, randomized, controlled trial.}
\newblock {\em medRxiv}, 2020.
\newblock URL:
  \url{https://www.medrxiv.org/content/early/2020/07/15/2020.07.15.20151852},
  \href
  {http://arxiv.org/abs/https://www.medrxiv.org/content/early/2020/07/15/2020.07.15.20151852.full.pdf}
  {\path{arXiv:https://www.medrxiv.org/content/early/2020/07/15/2020.07.15.20151852.full.pdf}},
  \href {http://dx.doi.org/10.1101/2020.07.15.20151852}
  {\path{doi:10.1101/2020.07.15.20151852}}.

\bibitem{ICH}
ICH.
\newblock {ICH E9 (R1) addendum on estimands and sensitivity analysis in
  clinical trials to the guideline on statistical principles for clinical
  trials}.
\newblock 2019.
\newblock URL:
  \url{https://www.ema.europa.eu/en/ich-e9-statistical-principles-clinical-trials}.

\bibitem{kang2007demystifying}
Joseph~DY Kang and Joseph~L Schafer.
\newblock Demystifying double robustness: A comparison of alternative
  strategies for estimating a population mean from incomplete data.
\newblock {\em Statistical Science}, 22(4):523--539, 2007.

\bibitem{Kon:2015}
Frank Konietschke, Arne~C Bathke, Solomon~W Harrar, and Markus Pauly.
\newblock Parametric and nonparametric bootstrap methods for general {MANOVA}.
\newblock {\em Journal of Multivariate Analysis}, 140:291--301, 2015.

\bibitem{Kunz2020}
Cornelia~Ursula Kunz, Silke J\"orgens, Frank Bretz, Nigel Stallard, Kelly~Van
  Lancker, Dong Xi, Sarah Zohar, Christoph Gerlinger, and Tim Friede.
\newblock Clinical trials impacted by the covid-19 pandemic: Adaptive designs
  to the rescue?
\newblock {\em Statistics in Biopharmaceutical Research}, 2020.
\newblock URL: \url{https://doi.org/10.1080/19466315.2020.1799857}, \href
  {http://arxiv.org/abs/https://doi.org/10.1080/19466315.2020.1799857}
  {\path{arXiv:https://doi.org/10.1080/19466315.2020.1799857}}, \href
  {http://dx.doi.org/10.1080/19466315.2020.1799857}
  {\path{doi:10.1080/19466315.2020.1799857}}.

\bibitem{Mao2020}
Huzhang Mao and Liang Li.
\newblock Flexible regression approach to propensity score analysis and its
  relationship with matching and weighting.
\newblock {\em Statistics in Medicine}, 2020.
\newblock \href {http://dx.doi.org/https://doi.org/10.1002/sim.8526}
  {\path{doi:https://doi.org/10.1002/sim.8526}}.

\bibitem{mao2019propensity}
Huzhang Mao, Liang Li, and Tom Greene.
\newblock Propensity score weighting analysis and treatment effect discovery.
\newblock {\em Statistical Methods in Medical Research}, 28(8):2439--2454,
  2019.

\bibitem{martinussen2013collapsibility}
Torben Martinussen and Stijn Vansteelandt.
\newblock {On collapsibility and confounding bias in Cox and Aalen regression
  models}.
\newblock {\em Lifetime Data Analysis}, 19(3):279--296, 2013.

\bibitem{McCaffrey2013}
Daniel~F McCaffrey, Beth~Ann Griffin, Daniel Almirall, Mary~Ellen Slaughter,
  Rajeev Ramchand, and Lane~F Burgette.
\newblock A tutorial on propensity score estimation for multiple treatments
  using generalized boosted models.
\newblock {\em Statistics in Medicine}, 32(19):3388--3414, 2013.

\bibitem{mccaw2020selecting}
Zachary~R McCaw, Lu~Tian, Kevin~N Sheth, Wan-Ting Hsu, W~Taylor Kimberly, and
  Lee-Jen Wei.
\newblock {Selecting appropriate endpoints for assessing treatment effects in
  comparative clinical studies for COVID-19}.
\newblock {\em Contemporary Clinical Trials}, 97:106145, 2020.
\newblock \href {http://dx.doi.org/10.1016/j.cct.2020.106145}
  {\path{doi:10.1016/j.cct.2020.106145}}.

\bibitem{muetze2020}
Tobias M{\"u}tze and Tim Friede.
\newblock {Data monitoring committees for clinical trials evaluating treatments
  of COVID-19}.
\newblock {\em Contemporary Clinical Trials}, 98:106154, 2020.
\newblock \href {http://dx.doi.org/10.1016/j.cct.2020.106154}
  {\path{doi:10.1016/j.cct.2020.106154}}.

\bibitem{PBK}
Markus Pauly, Edgar Brunner, and Frank Konietschke.
\newblock Asymptotic permutation tests in general factorial designs.
\newblock {\em Journal of the Royal Statistical Society: Series B (Statistical
  Methodology)}, 77(2):461--473, 2015.

\bibitem{pirracchio2012evaluation}
Romain Pirracchio, Matthieu Resche-Rigon, and Sylvie Chevret.
\newblock Evaluation of the propensity score methods for estimating marginal
  odds ratios in case of small sample size.
\newblock {\em BMC Medical Research Methodology}, 12(1):70, 2012.

\bibitem{puhr2017firth}
Rainer Puhr, Georg Heinze, Mariana Nold, Lara Lusa, and Angelika Geroldinger.
\newblock Firth's logistic regression with rare events: accurate effect
  estimates and predictions?
\newblock {\em Statistics in Medicine}, 36(14):2302--2317, 2017.

\bibitem{robins1986new}
James Robins.
\newblock A new approach to causal inference in mortality studies with a
  sustained exposure period -- application to control of the healthy worker
  survivor effect.
\newblock {\em Mathematical Modelling}, 7(9-12):1393--1512, 1986.

\bibitem{Robinson1991}
Laurence~D Robinson and Nicholas~P Jewell.
\newblock Some surprising results about covariate adjustment in logistic
  regression models.
\newblock {\em International Statistical Review/Revue Internationale de
  Statistique}, pages 227--240, 1991.

\bibitem{Rome2020}
Bejamnin~N. Rome and Jerry Avorn.
\newblock Drug evaluation during the covid-19 pandemic.
\newblock {\em New England Journal of Medicine}, 382(24):2282--2284, 2020.

\bibitem{Rosenbaum1983}
Paul~R Rosenbaum and Donald~B Rubin.
\newblock The central role of the propensity score in observational studies for
  causal effects.
\newblock {\em Biometrika}, 70(1):41--55, 1983.

\bibitem{Rosenbaum1984}
Paul~R Rosenbaum and Donald~B Rubin.
\newblock Reducing bias in observational studies using subclassification on the
  propensity score.
\newblock {\em Journal of the American Statistical Association},
  79(387):516--524, 1984.

\bibitem{rubin2008objective}
Donald~B Rubin.
\newblock For objective causal inference, design trumps analysis.
\newblock {\em The Annals of Applied Statistics}, 2(3):808--840, 2008.

\bibitem{rubin1996matching}
Donald~B Rubin and Neal Thomas.
\newblock Matching using estimated propensity scores: relating theory to
  practice.
\newblock {\em Biometrics}, pages 249--264, 1996.

\bibitem{Sattui2020}
Sebastian~E. Sattui, Jean~W. Liew, Elizabeth~R. Graef, Ariella Coler-Reilly,
  Francis Berenbaum, Ali Duarte-Garcia, Carly Harrison, Maximilian~F. Konig,
  Peter Korsten, Michael~S. Putman, Philip~C. Robinson, Emily Sirotich,
  Manuel~F. Ugarte-Gil, Kate Webb, Kristen~J. Young, Alfred~H.J. Kim, and
  Jeffrey~A. Sparks.
\newblock {Swinging the pendulum: lessons learned from public discourse
  concerning hydroxychloroquine and COVID-19}.
\newblock {\em Expert Review of Clinical Immunology}, 2020.
\newblock \href
  {http://dx.doi.org/https://doi.org/10.1080/1744666X.2020.1792778}
  {\path{doi:https://doi.org/10.1080/1744666X.2020.1792778}}.

\bibitem{Sjoelander2016}
Arvid Sj{\"o}lander, Elisabeth Dahlqwist, and Johan Zetterqvist.
\newblock A note on the noncollapsibility of rate differences and rate ratios.
\newblock {\em Epidemiology}, 27(3):356--359, 2016.

\bibitem{snowden2011implementation}
Jonathan~M Snowden, Sherri Rose, and Kathleen~M Mortimer.
\newblock {Implementation of G-computation on a simulated data set:
  demonstration of a causal inference technique}.
\newblock {\em American Journal of Epidemiology}, 173(7):731--738, 2011.

\bibitem{sonis}
Jeffrey Sonis.
\newblock {Odds Ratios vs Risk Ratios}.
\newblock {\em JAMA}, 320(19):2041--2041, 2018.
\newblock URL: \url{https://doi.org/10.1001/jama.2018.14417}, \href
  {http://arxiv.org/abs/https://jamanetwork.com/journals/jama/articlepdf/2715584/jama\_sonis\_2018\_le\_180130.pdf}
  {\path{arXiv:https://jamanetwork.com/journals/jama/articlepdf/2715584/jama\_sonis\_2018\_le\_180130.pdf}},
  \href {http://dx.doi.org/10.1001/jama.2018.14417}
  {\path{doi:10.1001/jama.2018.14417}}.

\bibitem{Stallard2020}
Nigel Stallard, Lisa Hampson, Norbert Benda, Werner Brannath, Thomas Burnett,
  Tim Friede, Peter~K. Kimani, Franz Koenig, Johannes Krisam, Pavel Mozgunov,
  Martin Posch, James Wason, Gernot Wassmer, John Whitehead, S.~Faye
  Williamson, Sarah Zohar, and Thomas Jaki.
\newblock {Efficient adaptive designs for clinical trials of interventions for
  COVID-19}.
\newblock {\em Statistics in Biopharmaceutical Research}, 2020.
\newblock \href {http://arxiv.org/abs/arXiv:2005.13309v1}
  {\path{arXiv:arXiv:2005.13309v1}}.

\bibitem{steyerberg2019clinical}
Ewout~W Steyerberg.
\newblock {\em Clinical prediction models}.
\newblock Springer, 2019.

\bibitem{stitelman2012general}
Ori~M Stitelman, Victor De~Gruttola, and Mark~J van~der Laan.
\newblock A general implementation of tmle for longitudinal data applied to
  causal inference in survival analysis.
\newblock {\em The International Journal of Biostatistics}, 8(1), 2012.

\bibitem{van2010collaborative}
Mark~J van~der Laan and Susan Gruber.
\newblock Collaborative double robust targeted maximum likelihood estimation.
\newblock {\em The International Journal of Biostatistics}, 6(1), 2010.

\bibitem{van2016no}
Maarten van Smeden, Joris~AH de~Groot, Karel~GM Moons, Gary~S Collins,
  Douglas~G Altman, Marinus~JC Eijkemans, and Johannes~B Reitsma.
\newblock No rationale for 1 variable per 10 events criterion for binary
  logistic regression analysis.
\newblock {\em BMC Medical Research Methodology}, 16(1):163, 2016.

\bibitem{van2019sample}
Maarten van Smeden, Karel~GM Moons, Joris~AH de~Groot, Gary~S Collins,
  Douglas~G Altman, Marinus~JC Eijkemans, and Johannes~B Reitsma.
\newblock Sample size for binary logistic prediction models: beyond events per
  variable criteria.
\newblock {\em Statistical Methods in Medical Research}, 28(8):2455--2474,
  2019.

\bibitem{wang2020}
Jixian Wang.
\newblock To use or not to use propensity score matching?
\newblock {\em Pharmaceutical Statistics}, 2020.
\newblock \href {http://dx.doi.org/https://doi.org/10.1002/pst.2051}
  {\path{doi:https://doi.org/10.1002/pst.2051}}.

\bibitem{wang2017flame}
Tianyu Wang, Marco Morucci, M~Awan, Yameng Liu, Sudeepa Roy, Cynthia Rudin, and
  Alexander Volfovsky.
\newblock {Flame: A fast large-scale almost matching exactly approach to causal
  inference}.
\newblock {\em arXiv preprint arXiv:1707.06315}, 2017.

\bibitem{Weitzen_2004}
Sherry Weitzen, Kate~L. Lapane, Alicia~Y. Toledano, Anne~L. Hume, and Vincent
  Mor.
\newblock {Weaknesses of goodness-of-fit tests for evaluating propensity score
  models: the case of the omitted confounder}.
\newblock {\em Pharmacoepidemiology and Drug Safety}, 14(4):227--238, jul 2004.
\newblock URL: \url{https://doi.org/10.1002%2Fpds.986}, \href
  {http://dx.doi.org/10.1002/pds.986} {\path{doi:10.1002/pds.986}}.

\end{thebibliography}


\begin{thebibliography}{1}

\bibitem{austin2008critical}
Peter~C Austin.
\newblock A critical appraisal of propensity-score matching in the medical
  literature between 1996 and 2003.
\newblock {\em Statistics in Medicine}, 27(12):2037--2049, 2008.

\bibitem{Baek2015}
Seunghee Baek, Seong~Ho Park, Eugene Won, Yu~Rang Park, and Hwa~Jung Kim.
\newblock Propensity score matching: a conceptual review for radiology
  researchers.
\newblock {\em Korean Journal of Radiology}, 16(2):286--296, 2015.

\bibitem{Elze2017}
Markus~C Elze, John Gregson, Usman Baber, Elizabeth Williamson, Samantha
  Sartori, Roxana Mehran, Melissa Nichols, Gregg~W Stone, and Stuart~J Pocock.
\newblock Comparison of propensity score methods and covariate adjustment:
  evaluation in 4 cardiovascular studies.
\newblock {\em Journal of the American College of Cardiology}, 69(3):345--357,
  2017.

\bibitem{Stuart2008}
Elizabeth~A. Stuart.
\newblock {Developing practical recommendations for the use of propensity
  scores: Discussion of 'A critical appraisal of propensity score matching in
  the medical literature between 1996 and 2003' by Peter Austin, Statistics in
  Medicine}.
\newblock {\em Statistics in Medicine}, 27(12):2062--2065, 2008.
\newblock URL: \url{https://onlinelibrary.wiley.com/doi/abs/10.1002/sim.3207},
  \href
  {http://arxiv.org/abs/https://onlinelibrary.wiley.com/doi/pdf/10.1002/sim.3207}
  {\path{arXiv:https://onlinelibrary.wiley.com/doi/pdf/10.1002/sim.3207}},
  \href {http://dx.doi.org/10.1002/sim.3207} {\path{doi:10.1002/sim.3207}}.

\end{thebibliography}

\end{document}

\title{\Large \bf Supplement to Causal inference methods for small non-randomized studies: Methods and an application in COVID-19
}
\author{Sarah Friedrich and  Tim Friede \\[1ex] 
{\small Department of Medical Statistics, University Medical Center G\"ottingen, Germany\footnote{Humboldtallee 32, 37073 G\"ottingen, Germany}}\\
{\small sarah.friedrich@med.uni-goettingen.de, tim.friede@med.uni-goettingen.de}
}
\maketitle

\begin{abstract} 
	%(max. 250 words)
In this supplemental material to our paper " Causal inference methods for small non-randomized studies: Methods and an application in COVID-19", we present the \redd{simulation results for the odds ratio}.
\end{abstract}

\newpage

\section{Simulation results for the Odds Ratio}

\redd{The simulation settings and the methods applied are the same as described for the risk difference in the paper. The following changes have to be applied to the methods in order to estimate the odds ratio instead of the risk difference:
	\begin{enumerate}
		\item The crude OR is estimated using logistic regression with treatment allocation as the only covariate.
		\item Covariate adjustment is performed for the logistic regression model, similarly also for the PS covariate method.
		\item When it comes to analyzing the matched data set, recommendations as to whether a matched-pair analysis is required or not are not entirely clear, see e.~g.~\cite{austin2008critical,Baek2015,Elze2017,Stuart2008} for discussions of this point.
Thus, we compared three different approaches to analyze the matched data set:
\begin{enumerate}
	\item {\bf Match unadjusted:} A logistic regression model for the outcome conditional on treatment exposure was implemented. This method does not account for the matched pairs.
	\item {\bf Match conditional:} A conditional logistic regression model accounting for the matched pairs was implemented. This is achieved by the \emph{clogit} function in R.
	\item {\bf Match GEE:} The logistic regression model was estimated using generalized estimating equations (GEE), which allows for specification of the matched pairs and estimation of robust standard errors. This approach was implemented using the \emph{geepack}-package in R. There are different ways to specify the correlation structure, e.~g.~using an exchangeable correlation matrix.
\end{enumerate}
\item The IPTW population is analyzed using weighted logistic regression with robust standard errors obtained from the \emph{sandwich}-package in R.
\item The g-computation approaches work exactly as described before, we only need to calculate the causal OR instead of the risk difference in the last step.
\end{enumerate}
}

\begin{table}[ht]
	\centering
	\caption{Overview of the simulated scenarios}
	\label{tab:overview}
	\begin{tabular}{c|cc||ccc}
		& $\beta_{trt}$ & true marginal OR & $\beta_0$ & \makecell{Percent treated\\ on average} & \makecell{simulated \\for true OR} \\ \hline
		& 0 & 1  &$-2.3$ & 55.2\% & 1, 2, 10 \\
		Scenario 1	& 0.8678 & 2 & $-1.8$ & 65.6\% & 1 \\
		& 2.7565 & 10    &$-1$ & 79.7\% & 1   \\ \hline
		& 0 & 1 &  & &\\
		Scenario 2  & 1.1111 & 2   & $-2.3$ & 53.8\% & 1, 2, 10 \\
		& 3.4793 & 10 &   & & \\ \hline
		& 0 & 1   & $-3.5$ & 49.4\% & 1, 2, 10 \\
		Scenario 3 & 0.9707 & 2 & $-1.5$& 80.1\% & 1 \\
		& 3.2625 & 10  & && \\ \hline
	%	Results: & \multicolumn{2}{c||}{ Figures~\ref{fig:bias}--\ref{fig:cov}, Tables~\ref{tab:Scen1}--\ref{tab:Scen3}}& \multicolumn{3}{c}{Figures~\ref{fig:Scen1_unb} and \ref{fig:Scen3_unb}}
	\end{tabular}
\end{table}

\begin{figure}[ht]
	\includegraphics[width = \textwidth]{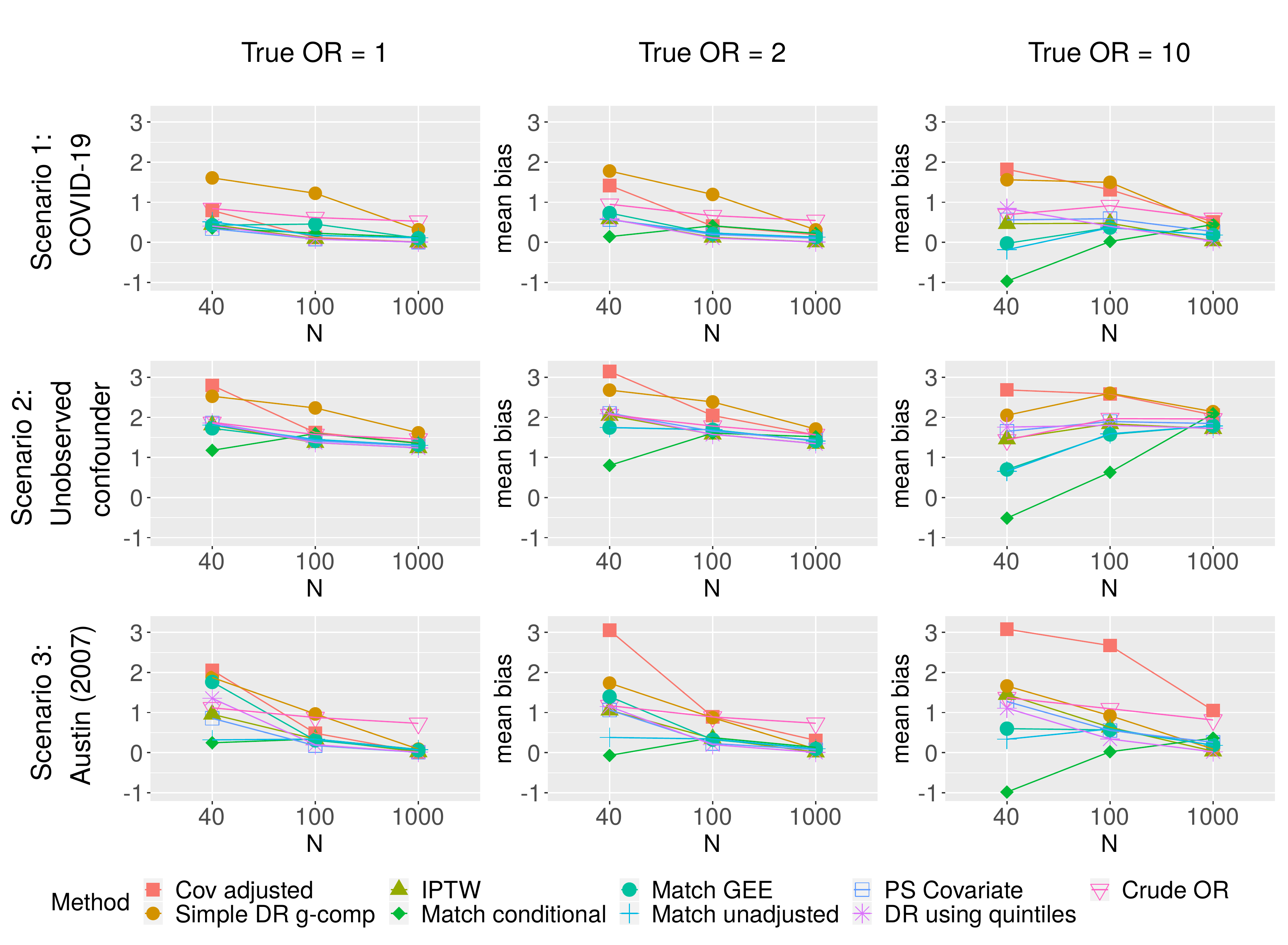}
	\caption{Displayed is the mean bias on the log-scale for the three scenarios (rows) and the three simulated marginal odds ratios (columns).}
	\label{fig:bias}
\end{figure}

\begin{figure}[ht]
	\includegraphics[width = \textwidth]{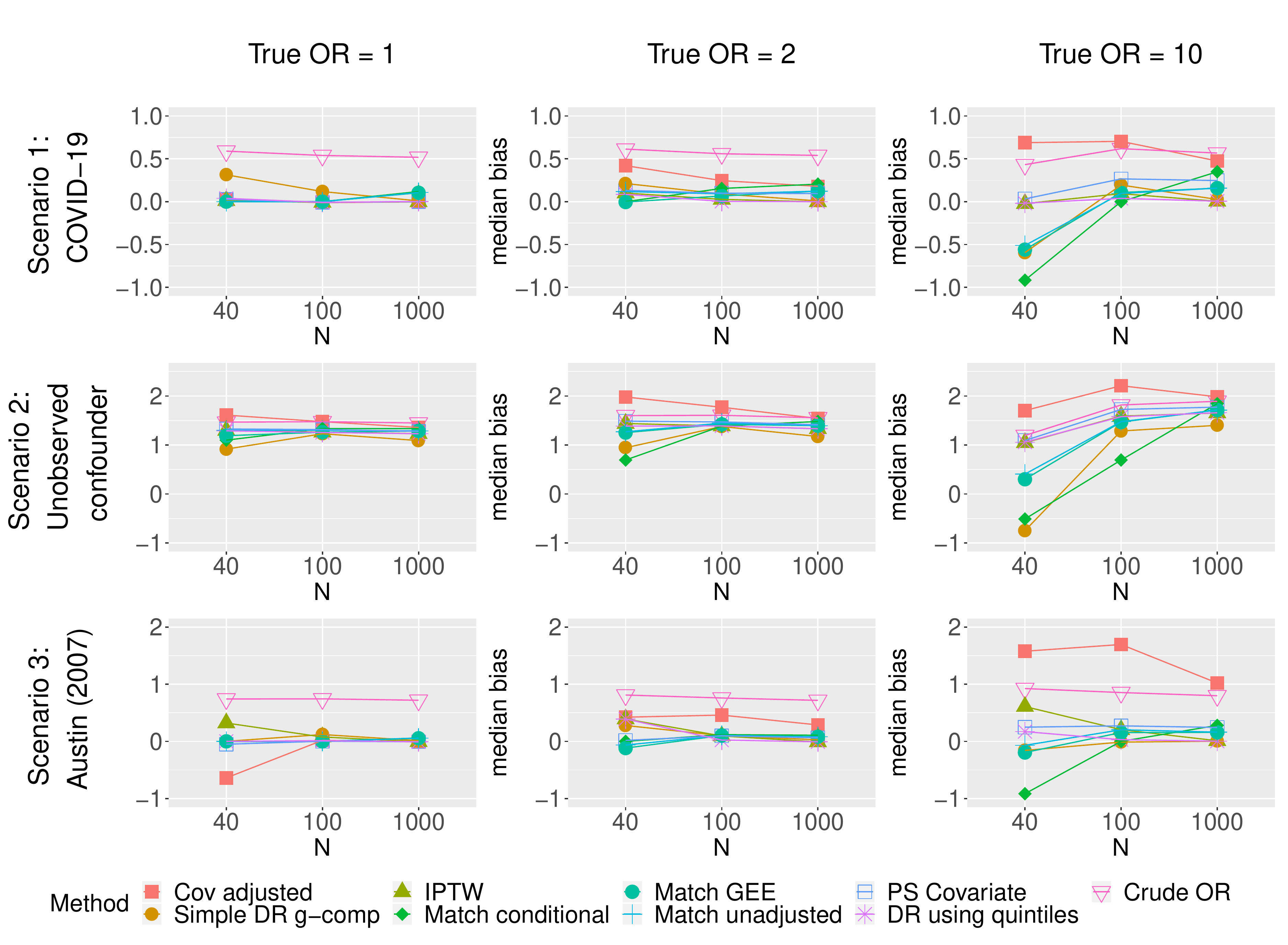}
	\caption{Displayed is the median bias on the log-scale for the three scenarios (rows) and the three simulated marginal odds ratios (columns).}
	\label{fig:medianbias}
\end{figure}

We used different measures to compare the results. With respect to the point estimators, we considered the bias on the log-scale, i.~e.~the difference between the true marginal log OR and the estimated log OR. Since the methods often resulted in extreme estimators of the treatment effect, we considered both the mean bias (difference between true OR and mean estimated treatment effect) and the median bias (difference between true OR and the median of the estimated treatment effect). The results are displayed in Figures~\ref{fig:bias} and \ref{fig:medianbias}, respectively. Moreover, the \red{root} mean square error of each estimated marginal OR (RMSE) is displayed in Tables~\ref{tab:Scen1}--\ref{tab:Scen3}.

\begin{figure}[ht]
	\includegraphics[width = \textwidth]{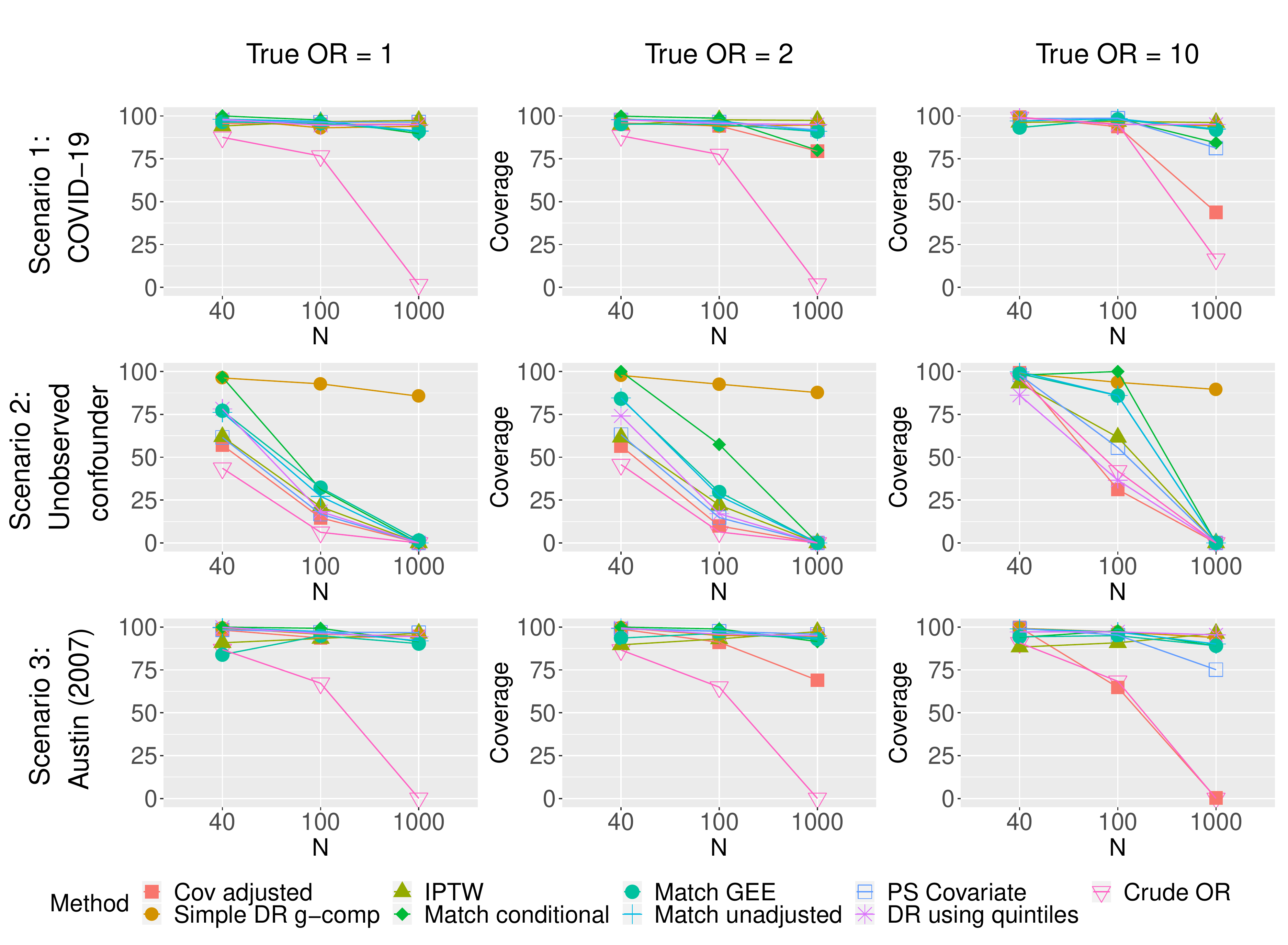}
	\caption{Displayed is the coverage probability (in \%) for the three scenarios (rows) and the three simulated marginal odds ratios (columns).}
	\label{fig:cov}
\end{figure}

Concerning the confidence intervals, we considered the percentage of 95\% confidence intervals that contained the true odds ratio (coverage probability) as well as the median length of the 95\% confidence interval. These measures are displayed in Figure~\ref{fig:cov} and  Tables~\ref{tab:Scen1}--\ref{tab:Scen3}, respectively. 
Finally, we also reported how often the chosen models did not converge or yielded an estimated OR $\geq 3000$. These were excluded from the calculations and reported as failures in the tables.

\begin{sidewaystable}[ht]
	\centering
	\caption{Median length of the 95\% confidence intervals (Length CI), \red{root} mean square error of the estimated treatment effect (\red{RMSE}) and number of models (out of 2000 simulation runs) that did not converge or resulted in an OR $> 3000$ for Scenario 1: COVID-19. Crude OR refers to the simple logistic regression adjusting only for treatment assignment, \red{Cov adjustment additionally adjusts for baseline covariates,} PS covariate denotes the method including the PS in the outcome regression model and \red{Simple DR g-comp and DR using quintiles} refer to the doubly robust g-computation method\red{s, respectively}.}
	\label{tab:Scen1}
	\begin{tabular}{cccccccccccc}
		\hline
		True OR & N& 	& \makecell{Crude \\ OR} & \makecell{\red{Cov}\\ \red{adjusted}}& \makecell{PS \\Covariate } & IPTW & \makecell{Simple DR \\g-comp} & \makecell{\red{DR using} \\ \red{quintiles}} & \makecell{Match \\ unadjusted} & \makecell{Match \\ conditional} & \makecell{Match \\ GEE}\\% & N & trueOR \\ 
		\hline
			
			\multirow{3}{*}{1} & \multirow{3}{*}{$N=40$}& CI Length & 5.912 & 6.037 & 4.519 & 4.195 & 28.36 & 8.301 & 6.736 & 6.958 & 5.998 \\ 
		& & RMSE & 2.642 & 7.063 & 1.527 & 2.031 & 9.059 & 1.817 & 2.094 & 1.23 & 1.946 \\ 
		& & Failures & 2 & 6 & 2 & 2 & 1 & 2 & 18 & 337 & 113 \\ 
			\hline
		& \multirow{3}{*}{$N=100$} & CI Length & 3.05 & 2.336 & 2.116 & 2.044 & 15.54 & 1.925 & 2.519 & 2.737 & 2.486 \\ 
		& & RMSE & 1.178 & 0.6674 & 0.5254 & 0.5315 & 5.158 & 0.487 & 0.6387 & 0.8687 & 18.31 \\ 
		& & Failures & 0 & 0 & 0 & 0 & 0 & 0 & 0 & 1 & 51 \\ 
		\hline
		& \multirow{3}{*}{$N=1000$} & CI Length & 0.8548 & 0.6052 & 0.5757 & 0.5546 & 4.016 & 0.4826 & 0.6565 & 0.6999 & 0.6563 \\ 
		& & RMSE & 0.7283 & 0.1534 & 0.139 & 0.1246 & 1.192 & 0.1223 & 0.1937 & 0.2207 & 0.2165 \\ 
		& & Failures & 0 & 0 & 0 & 0 & 0 & 0 & 0 & 0 & 11 \\ 
			\hline \hline
		\multirow{3}{*}{2} & \multirow{3}{*}{$N=40$} & CI Length & 14.14 & 20.65 & 11.1 & 10.33 & 75.63 & 23.22 & 15.2 & 14.26 & 13.87 \\ 
		& & RMSE & 6.15 & 30.99 & 6.731 & 5.546 & 60.98 & 6.727 & 4.516 & 1.57 & 41.26 \\ 
		& & Failures & 4 & 37 & 4 & 4 & 3 & 3 & 80 & 485 & 189 \\ 
			\hline
		& \multirow{3}{*}{$N=100$} & CI Length & 6.713 & 6.411 & 4.863 & 4.55 & 31.41 & 4.515 & 5.604 & 6.613 & 5.548 \\ 
		& & RMSE & 2.668 & 2.12 & 1.306 & 1.163 & 10.23 & 1.08 & 1.527 & 2.611 & 1.614 \\ 
		& & Failures & 0 & 0 & 0 & 0 & 0 & 0 & 0 & 29 & 64 \\ 
			\hline
		& \multirow{3}{*}{$N=1000$}  & CI Length & 1.863 & 1.505 & 1.299 & 1.184 & 8.301 & 1.035 & 1.399 & 1.693 & 1.406 \\ 
		& & RMSE & 1.531 & 0.5763 & 0.3875 & 0.2736 & 2.406 & 0.2676 & 0.4305 & 0.6615 & 0.437 \\ 
		& & Failures & 0 & 0 & 0 & 0 & 0 & 0 & 0 & 0 & 2 \\ 
			\hline \hline
		\multirow{3}{*}{10} & \multirow{3}{*}{$N=40$} & CI Length & 113 & 304.6 & 91.59 & 77.36 & 8.811$\cdot 10^{11}$ & 2.779$\cdot 10^{11}$ & 67.11 & 35.34 & 56.45 \\ 
		& & RMSE & 21.43 & 204 & 55.25 & 26.56 & 149.1 & 93.19 & 7.556 & 6.444 & 56.07 \\ 
		& & Failures & 521 & 822 & 522 & 521 & 373 & 323 & 927 & 1425 & 994 \\ 
		\hline
		& \multirow{3}{*}{$N=100$} & CI Length & 62.98 & 91.52 & 46.12 & 40.95 & 23153036 & 408792930 & 47.75 & 76.84 & 47.38 \\ 
		& & RMSE & 24.55 & 68.07 & 17.51 & 17.41 & 82.79 & 16.95 & 11.36 & 4.545 & 11.58 \\ 
		& & Failures & 74 & 92 & 74 & 74 & 69 & 71 & 197 & 701 & 253 \\ 
		\hline
		& \multirow{3}{*}{$N=1000$} & CI Length & 14.42 & 14.76 & 10.8 & 8.993 & 54.91 & 8.515 & 10.62 & 18.07 & 10.66 \\ 
		& & RMSE & 9.004 & 7.719 & 4.191 & 2.297 & 15.93 & 2.146 & 3.359 & 8.592 & 3.404 \\ 
		& & Failures & 0 & 0 & 0 & 0 & 0 & 0 & 0 & 0 & 1 \\ 
		\hline
		
	\end{tabular}
	
\end{sidewaystable}

\begin{sidewaystable}[ht]
	\centering
	\caption{Median length of the 95\% confidence intervals (Length CI), \red{root} mean square error of the estimated treatment effect (\red{RMSE}) and number of models (out of 2000 simulation runs) that did not converge or resulted in an OR $> 3000$ for Scenario 2: Unmeasured Confounder. Crude OR refers to the simple logistic regression adjusting only for treatment assignment, \red{Cov adjustment additionally adjusts for baseline covariates,} PS covariate denotes the method including the PS in the outcome regression model and \red{Simple DR g-comp and DR using quintiles} refer to the doubly robust g-computation method\red{s, respectively}.} 
	\label{tab:Scen2}
	\begin{tabular}{cccccccccccc}
		\hline
		True OR & N& 	& \makecell{Crude \\ OR} & \makecell{\red{Cov}\\ \red{adjusted}}& \makecell{PS \\Covariate } & IPTW & \makecell{Simple DR \\g-comp} & \makecell{\red{DR using} \\ \red{quintiles}} & \makecell{Match \\ unadjusted} & \makecell{Match \\ conditional} & \makecell{Match \\ GEE}\\% & N & trueOR \\ 
		\hline

			\multirow{3}{*}{1} & \multirow{3}{*}{$N=40$}  & CI Length & 15.93 & 27.6 & 16.42 & 15.98 & 80.72 & 32.48 & 21.75 & 19.85 & 22.99 \\ 
		& & RMSE & 10.09 & 71.51 & 17.54 & 11.29 & 25.56 & 21.93 & 9.23 & 2.973 & 8.931 \\ 
		& & Failures & 6 & 32 & 6 & 6 & 6 & 4 & 57 & 545 & 183 \\ 
		\hline
		& \multirow{3}{*}{$N=100$}   & CI Length & 8.405 & 9.831 & 7.646 & 7.422 & 39.2 & 8.072 & 8.786 & 10.63 & 8.854 \\ 
		& & RMSE & 4.421 & 5.217 & 3.698 & 3.506 & 14.35 & 3.549 & 4.127 & 5.318 & 4.247 \\ 
		& & Failures & 0 & 0 & 0 & 0 & 0 & 0 & 0 & 31 & 93 \\ 
		\hline
		& \multirow{3}{*}{$N=1000$}  & CI Length & 2.285 & 2.238 & 1.991 & 1.924 & 18.84 & 1.837 & 2.138 & 2.536 & 2.165 \\ 
		& & RMSE & 3.349 & 2.978 & 2.673 & 2.498 & 6.618 & 2.491 & 2.72 & 2.947 & 3.216 \\ 
		& & Failures & 0 & 0 & 0 & 0 & 0 & 0 & 0 & 0 & 35 \\ 
		
		\hline
		\hline
		\multirow{3}{*}{2} & \multirow{3}{*}{$N=40$} & CI Length & 45.63 & 108.5 & 46.45 & 43.65 & 350.8 & 157.1 & 56.03 & 35.34 & 54.46 \\ 
		& & RMSE & 24.21 & 164.5 & 39.76 & 28.38 & 67.02 & 50.88 & 15.74 & 3.249 & 28.49 \\ 
		& & Failures & 39 & 184 & 40 & 39 & 27 & 25 & 263 & 977 & 413 \\ 
		\hline
		&\multirow{3}{*}{$N=100$} & CI Length & 21.89 & 32.12 & 20.42 & 19.32 & 101.2 & 23.99 & 23.86 & 32.95 & 24.22 \\ 
		& & RMSE & 12.63 & 19.44 & 11.16 & 10.22 & 35.81 & 10.02 & 12.46 & 9.931 & 12.81 \\ 
		& & Failures & 0 & 0 & 0 & 0 & 0 & 0 & 1 & 248 & 76 \\ 
		\hline
		&\multirow{3}{*}{$N=1000$} & CI Length & 5.714 & 6.109 & 5.066 & 4.754 & 42.14 & 4.613 & 5.294 & 7.619 & 5.306 \\ 
		& & RMSE & 7.776 & 7.685 & 6.458 & 5.802 & 14.98 & 5.789 & 6.337 & 7.43 & 6.353 \\ 
		& & Failures & 0 & 0 & 0 & 0 & 0 & 0 & 0 & 0 & 7 \\ 
		
		\hline
		\hline		
		\multirow{3}{*}{10} & \multirow{3}{*}{$N=40$} & CI Length & 282.1 & 992.3 & 290.8 & 262.3 & 2.158$\cdot 10^{12}$ & 8.803$\cdot 10^{11}$ & 151 & 49.11 & 138.3 \\ 
		& & RMSE & 51.44 & 358.8 & 109 & 67.06 & 206.5 & 117.9 & 19.24 & 4.614 & 30.84 \\ 
		& & Failures & 1075 & 1322 & 1077 & 1075 & 709 & 716 & 1429 & 1761 & 1465 \\ 
		\hline
		&\multirow{3}{*}{$N=100$} & CI Length & 297.4 & 684.8 & 283.5 & 267.4 & 7.214$\cdot 10^{9}$ & 7.658$\cdot 10^{9}$ & 328.1 & 146.3 & 312.9 \\ 
		& & RMSE & 75.31 & 191 & 71.09 & 69.28 & 229.5 & 72.13 & 48.61 & 10.26 & 48.04 \\ 
		& & Failures & 404 & 456 & 404 & 404 & 370 & 402 & 657 & 1397 & 688 \\ 
		\hline
		&\multirow{3}{*}{$N=1000$} & CI Length & 75.46 & 88.63 & 67.68 & 61.92 & 352.3 & 69.45 & 69.38 & 146.3 & 69.88 \\ 
		& & RMSE & 65.44 & 73.68 & 57.32 & 49.5 & 128.7 & 49.6 & 53.98 & 89.12 & 54.24 \\ 
		& & Failures & 0 & 0 & 0 & 0 & 0 & 0 & 0 & 21 & 4 \\ 
		\hline
	\end{tabular}
	
\end{sidewaystable}

\begin{sidewaystable}[ht]
	\centering
	\caption{Median length of the 95\% confidence intervals (Length CI), \red{root} mean square error of the estimated treatment effect (\red{RMSE}) and number of models (out of 2000 simulation runs) that did not converge or resulted in an OR $> 3000$ for Scenario 3: Austin (2007). Crude OR refers to the simple logistic regression adjusting only for treatment assignment, \red{Cov adjustment additionally adjusts for baseline covariates,} PS covariate denotes the method including the PS in the outcome regression model and \red{Simple DR g-comp and DR using quintiles} refer to the doubly robust g-computation method\red{s, respectively}.} 
	\label{tab:Scen3}
	\begin{tabular}{cccccccccccc}
		\hline
		True OR & N& 	& \makecell{Crude \\ OR} & \makecell{\red{Cov}\\ \red{adjusted}}& \makecell{PS \\Covariate } & IPTW & \makecell{Simple DR \\g-comp} & \makecell{\red{DR using} \\ \red{quintiles}} & \makecell{Match \\ unadjusted} & \makecell{Match \\ conditional} & \makecell{Match \\ GEE}\\% & N & trueOR \\ 
		\hline
		
		\multirow{3}{*}{1} & \multirow{3}{*}{$N=40$} & CI Length & 9.028 & 74.72 & 9.211 & 7.77 & 52.76 & 23.66 & 19.31 & 15.92 & 6.996 \\ 
		& & RMSE & 3.69 & 49.91 & 14.4 & 4.14 & 14.73 & 48.33 & 1.784 & 0.9255 & 86.05 \\ 
		& & Failures & 56 & 340 & 66 & 56 & 29 & 13 & 266 & 925 & 322 \\ 
		\hline
		&\multirow{3}{*}{$N=100$} & CI Length & 4.504 & 4.129 & 3.022 & 3.23 & 16.76 & 4.05 & 3.928 & 3.878 & 3.919 \\ 
		& & RMSE & 1.938 & 5.32 & 0.8009 & 1.286 & 4.446 & 0.8193 & 1.263 & 1.22 & 1.358 \\ 
		& & Failures & 0 & 0 & 0 & 0 & 0 & 0 & 1 & 47 & 100 \\ 
		\hline
		&\multirow{3}{*}{$N=1000$}  & CI Length & 1.218 & 0.816 & 0.7266 & 0.8238 & 1.462 & 0.5933 & 0.8444 & 0.8581 & 0.8438 \\ 
		& & RMSE & 1.122 & 0.2211 & 0.173 & 0.1946 & 0.3965 & 0.1578 & 0.2422 & 0.2562 & 0.2825 \\ 
		& & Failures & 0 & 0 & 0 & 0 & 0 & 0 & 0 & 0 & 51 \\ 
		\hline 
		\hline
		\multirow{3}{*}{2} & \multirow{3}{*}{$N=40$} & CI Length & 18.26 & 113.4 & 17.15 & 16.41 & 89.43 & 42.94 & 22.13 & 15.92 & 17.96 \\ 
		& & RMSE & 8.013 & 221.2 & 69.15 & 10.11 & 25.14 & 40.6 & 3.338 & 1.164 & 91.49 \\ 
		& & Failures & 56 & 498 & 72 & 56 & 24 & 12 & 341 & 837 & 443 \\ 
			\hline
		&\multirow{3}{*}{$N=100$}  & CI Length & 8.968 & 12.08 & 6.018 & 6.701 & 27.4 & 6.984 & 7.24 & 8.334 & 7.347 \\ 
		& & RMSE & 3.904 & 7.728 & 1.566 & 2.502 & 8.264 & 1.692 & 2.336 & 2.326 & 2.342 \\ 
		& & Failures & 0 & 2 & 0 & 0 & 0 & 0 & 1 & 51 & 69 \\ 
			\hline
		&\multirow{3}{*}{$N=1000$}  & CI Length & 2.367 & 2.098 & 1.446 & 1.625 & 2.925 & 1.152 & 1.635 & 1.766 & 1.634 \\ 
		& & RMSE & 2.251 & 0.9054 & 0.3702 & 0.3783 & 0.7551 & 0.2978 & 0.4734 & 0.5724 & 0.483 \\ 
		& & Failures & 0 & 0 & 0 & 0 & 0 & 0 & 0 & 0 & 17 \\ 
			\hline
		\hline 
		\multirow{3}{*}{10} & \multirow{3}{*}{$N=40$} & CI Length & 150.2 & 2029 & 133.7 & 133.8 & 9.263$\cdot 10^{11}$ & 23044 & 111.7 & 35.34 & 111.7 \\ 
		& & RMSE & 53.34 & 502.5 & 123.3 & 83.43 & 131.5 & 74.38 & 14.69 & 6.481 & 74.72 \\ 
		& & Failures & 125 & 1377 & 177 & 125 & 57 & 28 & 671 & 1418 & 766 \\ 
			\hline
		&\multirow{3}{*}{$N=100$} & CI Length & 59.41 & 356.8 & 40.1 & 44.18 & 178.1 & 49.66 & 49.43 & 69.9 & 48.87 \\ 
		& & RMSE & 35.34 & 321.7 & 19.73 & 26.92 & 45.28 & 22.4 & 21.6 & 4.646 & 22.98 \\ 
		& & Failures & 1 & 87 & 1 & 1 & 1 & 1 & 7 & 651 & 183 \\ 
			\hline
		&\multirow{3}{*}{$N=1000$} & CI Length & 14.77 & 26.89 & 9.183 & 9.804 & 16.95 & 7.145 & 9.681 & 16.26 & 9.851 \\ 
		& & RMSE & 13.16 & 20 & 3.692 & 2.333 & 4.436 & 1.806 & 3.171 & 6.869 & 4.452 \\ 
		& & Failures & 0 & 0 & 0 & 0 & 0 & 0 & 0 & 0 & 68 \\ 
	\end{tabular}
	
\end{sidewaystable}

\begin{figure}[ht]
	\includegraphics[width=\textwidth]{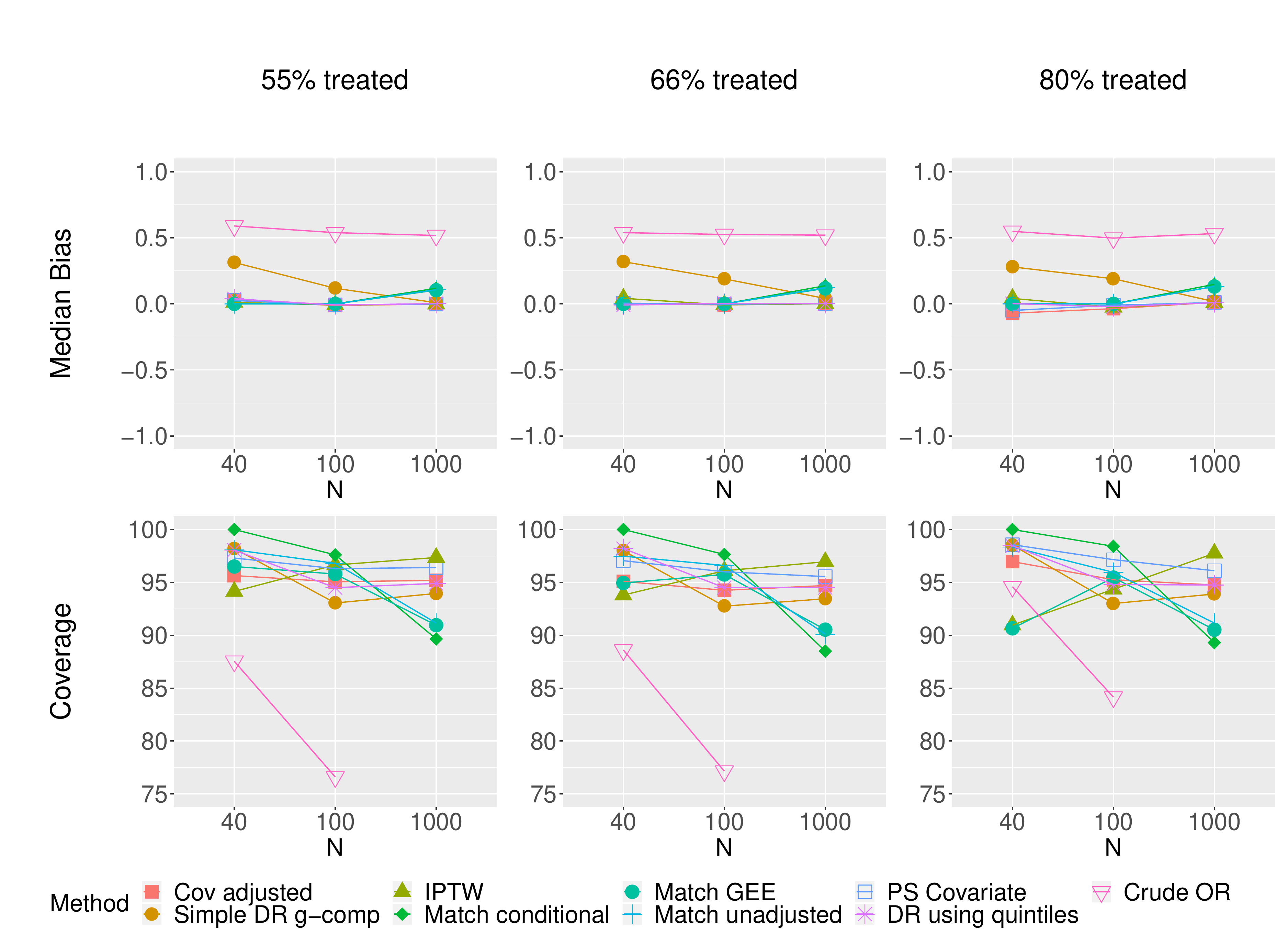}
	\caption{Median bias and coverage probabilities for Scenario 1 with a true OR of 1 and different proportions of treated individuals. Note that the coverage is truncated to $\geq 75\%$ implying that the unadjusted method is not displayed for $N=1000$.}
	\label{fig:Scen1_unb}
\end{figure}

\begin{figure}[ht]
	\centering
	\includegraphics[width=\textwidth]{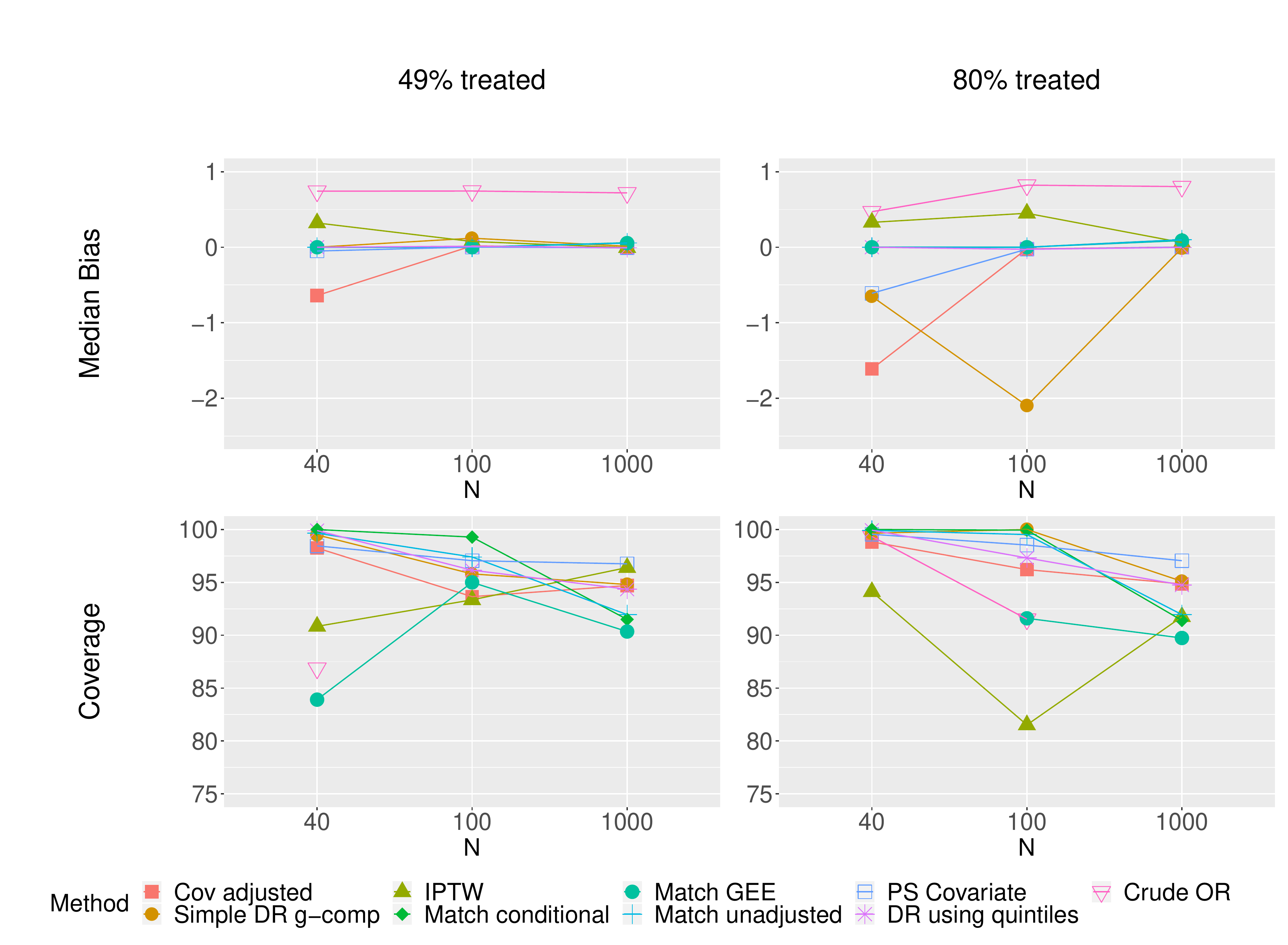}
	\caption{Median bias and coverage probabilities for Scenario 3 with a true OR of 1 and different proportions of treated individuals. Note that the coverage is truncated to $\geq 75\%$ implying that the unadjusted method is not displayed for some settings.}
	\label{fig:Scen3_unb}
\end{figure}

\bibliographystyle{plainurl}
\bibliography{Lit-Covid}